\documentstyle[12pt]{article}

\input FEYNMAN

\newcommand{\secn}[1]{Section~\ref{#1}}

\newcommand{\eq}[1]{Eq.~(\ref{#1})}
\newcommand{\fig}[1]{Fig.~\ref{#1}}
\newcommand{\nl}{\nonumber \\}

\def\beq{\begin{equation}}
\def\eeq{\end{equation}}
\def\beqa{\begin{eqnarray}}
\def\eeqa{\end{eqnarray}}

\newcommand{\sect}[1]{\setcounter{equation}{0}\section{#1}}
\renewcommand{\theequation}{\thesection.\arabic{equation}}
\newcommand{\EQ}{\begin{equation}}
\newcommand{\EN}{\end{equation}}
\newcommand{\bea}{\begin{eqnarray}}
\newcommand{\ena}{\end{eqnarray}}

\renewcommand{\a}{\alpha}

\newcommand{\e}{\epsilon}
\newcommand{\ve}{\varepsilon}

\renewcommand{\thefootnote}{\fnsymbol{footnote}}

\begin{document}
\begin{titlepage}
\rightline{DFTT 81/95}
\rightline{NORDITA 95/85-P}
\rightline{\hfill January 1996}
 
\vskip 1.2cm
 
\centerline{\Large \bf String techniques for the calculation} 
\centerline{\Large \bf of renormalization constants in field theory} 
 
\vskip 1.2cm
 
\centerline{\bf Paolo Di Vecchia\footnote{e-mail: DIVECCHIA@nbivax.nbi.dk}
 and Lorenzo Magnea\footnote{On leave from Universit\`a di Torino, Italy},}
\centerline{\sl NORDITA}
\centerline{\sl Blegdamsvej 17, DK-2100 Copenhagen \O, Denmark}
 
\vskip .2cm
  
\centerline{\bf Alberto Lerda\footnote{II Facolt\`a di Scienze 
M.F.N., Universit\`a di Torino 
(sede di Alessandria), Italy} and Rodolfo Russo,}
\centerline{\sl Dipartimento di Scienze e Tecnologie Avanzate and}
\centerline{\sl Dipartimento di Fisica Teorica, Universit\`a di Torino}
\centerline{\sl and I.N.F.N., Sezione di Torino}
\centerline{\sl Via P.Giuria 1, I-10125 Torino, Italy}

\vskip .2cm
 
\centerline{\bf Raffaele Marotta}
\centerline{\sl Dipartimento di Scienze Fisiche, Universit\`a di Napoli}
\centerline{\sl and I.N.F.N., Sezione di Napoli}
\centerline{\sl Mostra D'Oltremare, Pad. 19, I-80125 Napoli, Italy}

\vskip 1cm
 
\begin{abstract}
 
We describe a set of methods to calculate gauge theory renormalization 
constants from string theory, all based on a consistent prescription to 
continue off shell open bosonic string amplitudes.
We prove the consistency of our prescription by explicitly evaluating 
the renormalizations of the two, three and four-gluon amplitudes, and 
showing that they obey the appropriate Ward identities. 
The field theory limit thus performed corresponds to the 
background field method in Feynman gauge. 
We identify precisely the regions in string moduli space
that correspond to different classes of Feynman diagrams, and in particular
we show how to isolate contributions to the effective action.  
Ultraviolet divergent terms are then encoded in a single string integral
over the modular parameter $\tau$.
Finally, we derive a multiloop expression for the 
effective action by computing the partition function of an open bosonic 
string interacting with an external non-abelian background gauge field. 

\end{abstract}

\end{titlepage}

\newpage
\renewcommand{\thefootnote}{\arabic{footnote}}
\setcounter{footnote}{0}
\setcounter{page}{1}
  
\sect{Introduction}
\label{intro}
\vskip 0.5cm

String theories, unlike field theories, contain a physical dimensional 
parameter, the Regge slope $\alpha '$, that acts as an 
ultraviolet cutoff in 
the integrals over loop momenta, making multiloop amplitudes free from 
ultraviolet divergences. This is a basic reason why a 
string theory can
provide a consistent perturbative quantum theory of gravity, unified with 
non-abelian gauge theories.

Another useful feature of string theory is the fact that, 
at each order of string perturbation theory, one does not get 
the large proliferation of diagrams characteristic of field theories,
which makes it extremely difficult to perform high order calculations. 
In the case of closed strings one gets only one diagram at each order, 
while in the open string the number of diagrams remains small. 
Furthermore, the expression of these diagrams is known 
explicitly, in the case of the bosonic string, for an arbitrary 
perturbative order, including also the measure of integration on moduli 
space~\cite{copgroup}.

Finally, it is well known that in the limit of infinite string tension 
($\alpha ' \rightarrow 0$) string theories reduce to non-abelian gauge 
theories, unified with gravity, order by order in perturbation theory. 
This means, in particular, that in this limit one must reproduce, order 
by order, $S$-matrix elements, ultraviolet divergences,
and all other physical quantities that one computes in perturbative 
non-abelian gauge theories.

The combination of these different features of string 
theory has led
several authors~\cite{mets,mina,tayven,kaplu,berkosbet,berkoswfr} 
to show that
in many cases string theory can be an efficient conceptual 
and computational 
tool in different areas of perturbative field theory. 
In particular, because of the compactness of the multiloop string expression, 
it is easier to calculate non-abelian gauge theory amplitudes
by starting from a string theory, and performing the zero slope 
limit, rather than using traditional techniques. 
In this way the one-loop amplitude involving four external gluons has been 
computed, reproducing the known field theoretical result with much less 
computational cost~\cite{berkos}. Following the same approach, also 
the one-loop five-gluon amplitude has been computed for the first 
time~\cite{fiveglu}.

As observed in Ref.~\cite{berbos}, in order to extract 
field theory results 
from string theory, it is not 
necessary to use a consistent string theory. 
It can be sufficient to use the simplest string 
theory that contains the 
desired field theory in the zero slope limit (the open bosonic string if
one is interested in non-abelian gauge theories). String theory in fact 
proves itself to be a rather versatile and robust tool: as we 
shall see, one recovers
correct results in the field theory limit even when the consistency of the 
full string theory has been severely damaged, for example by continuing the
results to arbitrary space-time dimension, and by extending off shell the
string amplitudes, which implies a breaking of 
conformal invariance. Similar 
observations have led to the conjecture that the string organization may
be understood entirely in terms of a field theory analysis, and in fact
in some instances this has been shown~\cite{berdun,strass,schu}. However,
in view of a possible multiloop generalization of the known 
results~\cite{kaj}, we believe that further input from string theory 
may prove necessary.

In a previous letter~\cite{letter} we have considered the open bosonic 
string with Chan-Paton factors, and we 
have computed the one-loop two-gluon 
amplitude, with a suitable off-shell 
extrapolation in order to get rid of 
infrared divergences. By performing then the zero slope limit, we have 
obtained the gluon wave function renormalization constant $Z_A$, which 
turned out to coincide with the one 
calculated in the background field method.

In this paper we extend our procedure to 
the three and four-gluon amplitudes,
and thus compute also the renormalization 
constants $Z_3$, for the three-gluon
vertex, and $Z_4$, for the four-gluon vertex. This is a test of the 
consistency of our off-shell prescription, 
as we can show the validity of the 
Ward identities $Z_A = Z_3 = Z_4$, that are 
characteristic of the background 
field method. 

The information obtained with our method complements the
knowledge of how gauge-invariant on-shell scattering amplitudes can
be derived from string theory. In fact, in field theory, one is in general 
interested in calculating off-shell, gauge-dependent quantities such as
anomalous dimensions or general Green functions. An alternative set of rules
for calculating these objects is useful only if the choice of gauge,
regularization and renormalization prescriptions is uniquely specified.
The difficulty with string theory is of course that all one starts with is
a prescription to calculate scattering amplitudes, which is only valid
on shell. Different off-shell continuations must be shown to be consistent,
and may lead to different gauges or renormalization schemes. To date, the
only known consistent prescription for the off-shell continuation of string
amplitudes~\cite{berkosrol} leads to a vanishing wave-function renormalization
for the gauge field, in apparent contradiction with the results of a detailed
analysis of on-shell amplitudes~\cite{berdun}, which implies that string 
theory leads to a combination of the background field method with the 
non-linear Gervais-Neveu gauge. In the following we will solve this puzzle
by proposing a different, and simpler, prescription for the off-shell
continuation of the amplitudes, which leads unambiguously to the
background field method.

The calculation of renormalization constants can be performed using three
different, although closely related methods.
The most straightforward method, in the spirit of~\cite{berkosbet},
is based on the evaluation of the field theory limit of the full string 
amplitude, from which ultraviolet divergences can be consistently extracted 
after off-shell continuation. A careful analysis of the string amplitudes
leads to a precise identification of the regions in moduli space that 
correspond to different classes of Feynman diagrams in field theory.
Having learnt how to identify one-particle irreducible diagrams, contributing
to the effective action, one may look for a method to isolate these
contributions directly in the general expression for the amplitude. This is
achieved by using a different expression for the bosonic world-sheet
Green function, along the lines of~\cite{mets}. We establish a precise
relationship between these two methods, showing explicitly how the 
contributions to the effective action are picked out of the full string 
amplitude. Finally, having observed how string amplitudes seem to be 
intimately connected with the background field method, we go on to
show how effective actions with arbitrary numbers of legs and loops
can be compactly represented in terms of a generating functional
describing the propagation of an open string in an external non-abelian
gauge field. The renormalization constants derived from this
generating functional coincide with those previously obtained
calculating string scattering amplitudes. 

The first method (presented in Sections 3, 4 and 5) uses the open 
bosonic string amplitudes for two, three and four gluons, suitably 
continued off shell. The amplitudes are expressed as an 
integral over the puncture coordinates, representing 
the locations of the insertions of the gluon operators on the open 
string boundaries, and over the modular parameter
of the annulus, $\tau$. The integral over the punctures has a Laurent 
expansion in terms of the variable $k = e^{- 2 \tau}$. The 
leading term of this expansion, of order $k^{-1}$, corresponds to 
tachyon exchange in the loop, and must be discarded by hand. The 
next-to-leading term, of order $1$, corresponds to gluon exchange 
in the loop, and gives the field theory limit. 
Higher order contributions, behaving 
as $k^n$ with $n>0$, are 
negligible in the field theory limit. 
If we then concentrate on 
the term of order $1$, we see that the infinities
appearing in the field theory limit arise from different regions of 
integration over the punctures. There are basically two kinds of 
contributions~\cite{berkos}.  
The first one is obtained by expanding the integrand
for large $\tau$, isolating the term of order $1$ in $k$, and then
integrating the resulting expression over the punctures. This 
contribution corresponds 
to one-particle irreducible diagrams in field theory.
The second kind of contribution 
arises instead from regions of integration 
in which some punctures are taken to be very close to each other.
This is often referred to as the pinching limit, and corresponds to field
theory diagrams that are connected but 
not one-particle irreducible. 
The sum of the two contributions reproduces the infinities of the full 
connected Green functions, truncated with free propagators and suitably 
continued off shell. 

The second method is based on the use of an alternative form of the bosonic 
Green function, proposed in Ref.~\cite{fratse1}, in which the
logarithmic singularity associated with the pinching limit is
regularized by means of a $\zeta$-function regularization.
Using this regularized expression for the Green function
simplifies considerably the evaluation of renormalization constants.
In fact, as we shall see, having eliminated the reducible diagrams associated 
with the pinching regions, one can easily isolate ultraviolet divergent
contributions by focusing on the terms in the integrand with the appropriate 
power of the modular parameter. It is then possible to neglect the 
exponentials of the bosonic Green functions that constitute the bulk of 
the integrand. This simplifies the calculation to the point that it is
actually possible to perform the integrals over the punctures exactly,
before the field theory limit is taken. 

Finally, we show how all divergent contributions to the effective action 
can be generated from a single functional, describing the interaction of an
open bosonic string with an external non-abelian background gauge field.
Using this formalism, one sees the connection between string theory and
the background field method from a different angle. String-derived gluon
amplitudes without pinchings can be understood as scattering 
amplitudes in a background field constructed out of plane waves. 
Contributions to the
effective action can be isolated by focusing on slowly varying background
fields, and are then connected to gluon amplitudes for low energy gluons.
Manipulating the expression for the all-order partition function of the
string in such a slowly varying background, we can formally generalize
the calculation of the effective action, and we 
obtain a compact expression for the string integral generating the 
renormalization constants, valid for any number of string loops.   

The paper is organized as follows. In \secn{mglue} we consider the open 
bosonic string, and we write the explicit expression of the $M$-gluon 
amplitude at $h$ loops, including the overall normalization. We specialize 
then to the case of tree and one-loop diagrams, detailing various 
choices of integration variables.
In \secn{twopoint} we sketch  the calculation of 
the one-loop two-gluon 
amplitude, already presented in~\cite{letter}, and we show how to 
extract the gluon wave function renormalization constant $Z_A$. 
In \secn{threepoint} we extend our previous 
calculation 
to the one-loop three-gluon amplitude, and we discuss in detail the 
various regions of integration over the punctures that contribute to
the field theory limit. In particular, we show how one can identify at
the string level the classes of Feynman diagrams that correspond 
to each such region, and we derive the 
renormalization constant $Z_3$.
In \secn{fourpoint} we consider the one-loop four-gluon 
amplitude, we extract the 
renormalization constant $Z_4$, and we discuss some subtleties arising
in the treatment of contact terms. 
In \secn{effact} we show how, using a regularized expression of the Green 
function, one is naturally led to calculate contributions to the effective 
action, and finally in \secn{background} we recover and extend these 
results by computing the partition function of a string 
interacting with an external 
non-abelian field. \secn{summa} summarizes our results and compares the 
different approaches, having in mind possible applications to
multiloop calculations in gauge theories.

\sect{The $M$-gluon amplitude}
\label{mglue}
\vskip 0.5cm
 
The properly normalized (connected) $S$-matrix element describing the
scattering of $M$ string states $|\alpha_1\rangle,\dots,|\alpha_M\rangle$ 
with momenta $p_1,\dots,p_M$ is
\EQ
S(\alpha_1,\dots,\alpha_M) = {\rm i}\,(2\pi)^d\,\delta^{(d)}(p_1 + \dots
+ p_M)~A(\alpha_1,\dots,\alpha_M)\,\prod_{i=1}^M
\left(2E_i\right)^{-1/2}~~~,
\label{smatrix}
\EN
where $d$ is the dimension of the target space-time with metric 
$(-,+,\dots,+)$, $E_i=|p_i^0|$ is the energy of the $i$-th particle and 
$A(\alpha_1,\dots,\alpha_M)$ is the $M$-point amplitude in which 
all propagators of the $M$ external states 
have been removed. 
Borrowing the terminology from field theory, we call
$A(\alpha_1,\dots,\alpha_M)$ the (unrenormalized) connected $M$-point 
amplitude. In string theory, these amplitudes can be computed pertubatively, 
and are written as
\bea
A(\alpha_1,\dots,\alpha_M)&=&\sum_{h=0}^\infty
A^{(h)}(\alpha_1,\dots,\alpha_M) \nl
&=& \sum_{h=0}^\infty g_s^{2h-2}
{\hat A}^{(h)}(\alpha_1,\dots,\alpha_M)~~~,
\label{pertampl}
\ena
where $g_s$ is a dimensionless string coupling constant, which is
introduced to formally control the perturbative expansion.
In \eq{pertampl}, $A^{(h)}$ represents the $h$-loop contribution. 
In particular, for the open bosonic string, $A^{(0)}$
corresponds to the scattering of string states 
inserted on the 
boundary of a disk, while $A^{(1)}$ corresponds to the scattering of
states inserted either on one of the boundaries of an annulus (planar
diagram), or on the boundary of a Moebius 
strip (non-orientable diagram),
or on both boundaries of an annulus (non-planar diagram). Similarly
the higher loop amplitudes $A^{(h)}$ with $h\geq 2$ correspond to the 
scattering of states inserted on the various boundaries of an open Riemann 
surface with a more complicated topology. 
In what follows we will limit our discussion to planar 
diagrams, but our analysis can be extended to the other cases as well.
 
An efficient way to explicitly obtain $A^{(h)}(\alpha_1,\dots,\alpha_M)$ 
is to use the $M$-point $h$-loop vertex $V_{M;h}$ of the operator
formalism~\cite{copgroup}, which can be regarded as a generating functional 
for scattering amplitudes among arbitrary string states, at all orders in 
perturbation theory. In fact, by saturating the operator $V_{M;h}$  with
$M$ states $|\alpha_1\rangle, \ldots ,|\alpha_M\rangle$, one obtains the 
corresponding amplitude, 
\EQ
A^{(h)}(\alpha_1,\ldots,\alpha_M) =
V_{M;h}~ |\alpha_1\rangle \cdots |\alpha_M\rangle~~~.
\label{ampvert}
\EN
 
The explicit expression of $V_{M;h}$ for the planar diagrams of the
open bosonic string can be found in Ref.~\cite{copgroup}. That
expression still has to be properly normalized, and the correct
normalization factor $C_h$ is given by
\EQ
C_h = {1\over{(2\pi)^{dh}}}~g_s^{2h-2}{1\over{(2\a')^{d/2}}}~~~,
\label{vertnorm}
\EN
where $\a'$ is the inverse of the string tension (we refer to Appendix A 
for a derivation of this formula).
 
The vertex $V_{M;h}$ depends on $M$ real Koba-Nielsen variables $z_i$ 
through $M$ projective transformations $V_i(z)$, which define local 
coordinate systems vanishing around each $z_i$, {\it i.e.} such that 
\EQ
V_i^{-1}(z_i) = 0~~~.
\label{Vi}
\EN
When $V_{M;h}$ is saturated with $M$ physical string states satisfying 
the mass-shell condition, the corresponding 
amplitude does not depend on 
the $V_i$'s. However, as we discussed in Ref.~\cite{letter}, to extract
information about the ultraviolet divergences that arise when the field 
theory limit is taken, it is necessary to relax the mass-shell condition, so 
that also the amplitudes $A^{(h)}$ will depend on the choice of projective 
transformations $V_i$'s, just like the vertex $V_{M;h}$.
This is reminiscent of a well-known fact in gauge theories, namely that
on-shell amplitudes are gauge invariant, whilst their off-shell
counterparts are not. Indeed, one can regard the freedom of choosing
the projective transformations $V_i$'s as 
a sort of gauge freedom.
 
Let us now specialize to the case in which all $M$ states are gluons,
and let us denote by $|p,\ve\rangle$ a state representing a
gluon with momentum $p$ and polarization $\ve$. 
Such a state is created by the vertex operator
\beq
V(z) = {\rm i}\,{\cal N}_0
\,:\,\ve\cdot \partial_z X(z)~{\rm e}^{{\rm i}
\, \sqrt{2\a'}\,p\cdot X(z)}\,:~~~,
\label{gluvert}
\eeq
where colons denote the standard normal 
ordering on the modes of the open 
string coordinate $X(z)$, and ${\cal N}_0$ is a normalization factor 
given by
\beq
{\cal N}_0 = g_d \,\sqrt{2\a'}~~~,
\label{glunorm}
\eeq
with $g_d$ being the gauge coupling constant of the target space
Yang-Mills theory. The latter is related to the dimensionless 
string coupling constant $g_s$ of \eq{pertampl} by
\beq
g_s = \frac{g_d}{2}\,(2\a')^{1-d/4}~~~,
\label{gstring}
\eeq
as described in Appendix A.
Notice that the vertex in \eq{gluvert} does 
not depend on the color index 
of the gluon, and thus the gauge group structure 
must be recovered via the Chan-Paton procedure.
 
If we write as usual
\EQ
X^\mu(z) = {\hat q}^\mu - {\rm i}{\hat p}^\mu \log z + {\rm i}
\sum_{n\not=0} {{{\hat a}^\mu_n}\over n} z^{-n}~~~,
\label{coord}
\EN
then the gluon state is
\EQ
|p,\ve\rangle \equiv 
\lim_{z\to 0} V(z)|0\rangle =
{\cal N}_0 ~\ve \cdot {\hat a}_{-1} |p\rangle~~~,
\label{glustate}
\EN
where $|p\rangle$ is the ground state with momentum $p$.
The gluon is physical if 
\bea
p^2 &=& 0~~~,  \label{massshell} \\
\ve \cdot p &=& 0~~~. \label{transverse}
\ena 
We will always enforce the transversality condition (\ref{transverse}), 
while in discussing the field theory limit of the string 
amplitudes we will 
relax the mass-shell condition (\ref{massshell}) (see also~\cite{letter}). 
Relaxing also the transversality condition would 
just lead to more cumbersome 
expressions without changing the final results.
 
If we saturate the vertex $V_{M;h}$ with $M$ gluon states and multiply
the result with the appropriate Chan-Paton factor, we 
obtain the color-ordered $M$-gluon planar amplitude at $h$ loops,
which we denote by $A^{(h)}_P (p_1,\ldots,p_M)$. 
If we pick the gauge group $SU(N)$, use transverse 
gluons, but choose not to 
enforce the mass-shell condition, we find
\bea
A^{(h)}_P (p_1,\ldots,p_M) & = & N^h\,{\rm Tr}(\lambda^{a_1}
\cdots \lambda^{a_M})\,
C_h\,{\cal N}_0^M    \nl  
& \times & \int [dm]^M_h \left\{\prod_{i<j} 
\left[{{\exp\left({\cal G}^{(h)}(z_i,z_j)\right)}
\over{\sqrt{V'_i(0)\,V'_j(0)}}}\right]^{2\a' p_i\cdot p_j} \right.  \nl
& \times & \exp\left[\sum_{i\not=j}
\sqrt{2\a'}\, p_j\cdot\ve_i 
\,\partial_{z_i} {\cal G}^{(h)}(z_i,z_j) \right.  \label{hmaster}  \\
& & + \, \left.\left.{1\over 2}\sum_{i\not=j} 
\ve_i\cdot\ve_j \,\partial_{z_i}\partial_{z_j}
{\cal G}^{(h)}(z_i,z_j)\right]\right\}_{\rm m.l.}~~~,  \nonumber
\ena
where the subscript ``m.l.'' stands for multilinear, meaning
that only terms linear in each polarization 
should be kept. 
In \eq{hmaster}, which can be considered the master formula of
our approach, $N^h{\rm Tr}(\lambda^{a_1} \cdots \lambda^{a_M})$ 
is the Chan-Paton factor appropriate for an $M$-gluon $h$-loop planar 
diagram, with the $\lambda$'s being the generators of $SU(N)$ in the 
fundamental representation, normalized as
\beq
{\rm Tr}(\lambda^a\,\lambda^b) = \frac{1}{2}\,\delta^{ab}~~~.
\label{gennorm}
\eeq
Further, ${\cal G}^{(h)}(z_i,z_j)$ is the $h$-loop 
world-sheet bosonic 
Green function, and $[dm]^M_h$ is the measure of integration on moduli 
space for an open Riemann surface of genus $h$ with $M$ operator insertions
on the boundary~\cite{copgroup}.
The Green function ${\cal G}^{(h)}(z_i,z_j)$ can be expressed as
\beq
{\cal G}^{(h)}(z_i,z_j) = \log E^{(h)}(z_i,z_j) - {1\over 2} \int_{z_i}^{z_j} 
\omega^\mu \, \left(2\pi {\rm Im}\tau_{\mu\nu}\right)^{-1} 
\int_{z_i}^{z_j} \omega^\nu~~~, 
\label{hgreen}
\eeq
where $E^{(h)}(z_i,z_j)$ is the prime-form, $\omega^\mu$ ($\mu=1,\ldots, h$)
the abelian differentials and $\tau_{\mu\nu}$ the period
matrix of an open Riemann surface of genus $h$.
All these objects, as well as the measure on moduli space $[dm]^M_h$, can 
be explicitly written down in the Schottky parametrization of the Riemann 
surface, and their expressions for arbitrary $h$ can be found for example 
in~\cite{scho}. Here we will only reproduce the explicit
expression for the measure, to give a flavor of the ingredients that enter 
the full string theoretic calculations. It is
\beqa
[dm]^M_h & = & \frac{\prod_{i=1}^M dz_i}{dV_{abc}}
\prod_{\mu=1}^{h} \left[ \frac{dk_\mu d \xi_\mu d \eta_\mu}{k_\mu^2
(\xi_\mu - \eta_\mu)^2} ( 1- k_\mu )^2 \right]   \label{hmeasure} \\
& \times & \left[\det \left( - i \tau_{\mu \nu} \right) \right]^{-d/2} 
\prod_{\alpha}\;' \left[ \prod_{n=1}^{\infty} ( 1 - k_{\alpha}^{n})^{-d}
\prod_{n=2}^{\infty} ( 1 - k_{\alpha}^{n})^{2} \right]~~~.   \nonumber
\eeqa
Here $k_{\mu}$ are the 
multipliers and $\xi_{\mu}$ and $ \eta_{\mu}$ the fixed 
points of the generators of the Schottky group; $dV_{abc}$ 
is the projective 
invariant volume element 
\beq
dV_{abc} = \frac{d\rho_a~d\rho_b~d\rho_c}
{(\rho_a-\rho_b)~(\rho_a-\rho_c)~(\rho_b-\rho_c)}~~~,
\label{projvol}
\eeq
where $\rho_a$, $\rho_b$, $\rho_c$ are any three of the $M$
Koba-Nielsen variables, or of the $2h$ fixed points of the generators of the 
Schottky group, which can be fixed at will; finally, the primed product
over $\alpha$ denotes a product over classes of elements of the Schottky 
group~\cite{scho}.
Notice that in the open string the Koba-Nielsen variables must be cyclically 
ordered, for example according to
\beq 
z_1 \geq z_2 \geq \cdots \geq z_{M}~~~,
\label{cyclord}
\eeq
and the ordering of Koba-Nielsen variables automatically
prescribes the ordering of color indices. 

Clearly expressions such as \eq{hmeasure} can be very difficult to handle 
in practice, as they involve infinite products (or sums) over subgroups of a 
discrete group. However, as we shall see, all such expressions simplify
drastically in the field theory limit $\a' \rightarrow 0$. In this limit,
only a finite number of terms contribute to the products, and calculations
remain manageable. 

The reader may also have noticed that we have left the value of the space-time
dimension $d$ arbitrary, and we have not fixed it to the critical dimension,
$d = 26$. As we shall see, this flagrant violation of string consistency
does not affect the field theory limit. This may be understood if we 
interpret $d$ as the number of {\it uncompactified} dimensions, and we assume
that $d' = 26 - d$ dimensions have been compactified. Of course the
$d'$ compactified dimensions will 
contribute to the string partition function
and to the scattering amplitudes, but those contributions factorize, and, 
in the field theory limit, behave just as $d'$ scalars coupled to the gauge 
field. The na\"{\i}ve procedure, simply letting $d$ be arbitrary,
just corresponds to neglecting ab initio the contribution of these scalars.

For $h=0$ the situation is particularly simple. The Green
function in \eq{hgreen} reduces to 
\beq
{\cal G}^{(0)}(z_i,z_j) = \log (z_i-z_j)~~~,
\label{treegreen}
\eeq
while the measure $[dm]^M_0$ is simply
\beq
[dm]^M_0 = \frac{\prod\limits_{i=1}^M dz_i}{dV_{abc}}~~~.
\label{treemeas}
\eeq
Inserting \eq{treegreen} and \eq{treemeas} into \eq{hmaster}, 
and writing explicitly all the normalization coefficients, we obtain
the color-ordered, on-shell $M$-gluon amplitude 
at tree level\footnote{We omit the subscript $P$ for 
planar, since at tree level only planar 
diagrams are present.}
\bea
A^{(0)} (p_1,\ldots,p_M) & = & 4\,{\rm Tr}(\lambda^{a_1}
\cdots \lambda^{a_M})\,g_d^{M-2}\,(2\a')^{M/2-2}   \nl
& \times & \int_{\Gamma_0}\frac{\prod\limits_{i=1}^M dz_i}{dV_{abc}} 
\left\{\prod_{i<j} \left(z_i-z_j\right)^{2\a' p_i\cdot p_j} 
\right. \label{treemaster} \\
& \times & \left. \exp\left[\sum_{i<j}
\left(\sqrt{2\a'}\, \frac{p_j\cdot\ve_i 
-p_i\cdot\ve_j}{(z_i-z_j)} + \frac{\ve_i\cdot
\ve_j}{(z_i-z_j)^2}\right)\right]\right\}_{\rm m.l.}~~~, \nonumber
\ena
where $\Gamma_0$ is the region identified by \eq{cyclord}.
Notice that any dependence on the local coordinates $V_i(z)$
drops out in the amplitude after enforcing 
the mass-shell condition.
\eq{treemaster} is valid for $M \geq 3$, since the tree-level measure
\eq{treemeas} is ill-defined for $M \leq 2$.

For $h=1$ the situation is more complicated 
but still manageable.
In the Schottky representation, the annulus is described 
by a Schottky group generated by one real projective
transformation $S$. This can be characterized 
by a multiplier $k$ 
and the attractive and repulsive fixed points $\xi$ and $\eta$. 
In terms of these variables, the measure $[dm]^M_1$ on moduli space turns 
out to be
\beq
[dm]^M_1 = \frac{\prod_{i=1}^M dz_i}{dV_{abc}}~dk~d\xi~d\eta~
{1\over{k^2(\xi-\eta)^2}} \left(-{1\over {2\pi}}\ln k\right)^{-d/2}
\prod_{n=1}^\infty (1-k^n)^{2-d}~~~,
\label{onemeas}
\eeq
where $dV_{abc}$ is given again by \eq{projvol}, with $\rho_a$, $\rho_b$
and $\rho_c$ being any of the $M+2$ variables $\{z_1,\ldots,z_M,\xi,\eta\}$, 
which can be fixed at will. The remaining $M-1$ variables and $k$ are 
the moduli of the annulus with $M$ operator 
insertions on one boundary. 
In deriving \eq{onemeas} from \eq{hmeasure}, one should remember 
that the determinant of the period matrix, coming from the integration over 
the momenta circulating in the loops, can be expressed at one loop 
in terms of the multiplier of the Schottky generator as
\beq
k= e^{2\pi i {\tilde{\tau}}}~~~.
\label{ktau}
\eeq
Here, in the one-loop case, we have called the period matrix $\tilde{\tau}$ 
instead of $\tau$, as usually done, because later on we will use $\tau$ for a
rescaled version of it (see \eq{tau}).
 
A crucial point that must be discussed before we proceed is the region 
$\Gamma_1$ over which the moduli must be integrated. 
One way of determining 
$\Gamma_1$ is to observe that the vertex $V_{M;1}$ can be obtained with
the sewing procedure from $V_{M+2;0}$~\cite{DFLS}. As we mentioned before,
$V_{M+2;0}$ depends on $M+2$ Koba-Nielsen variables ordered on the real 
line, as in \eq{cyclord}.
In the sewing process one identifies, say, legs $M+1$ and $M+2$
by means of a propagator $P(x)$ depending on a sewing parameter $x$, 
with $0 < x < 1$. The specific form of $P(x)$ is 
determined by 
the local coordinates $V_i(z)$ of the two legs that are sewn, in 
such a way that the factorizability and the cyclic properties of the
vertex are maintained after sewing. We refer the reader to~\cite{DFLS}
for a thorough discussion of this point and of the sewing procedure 
in general.
Here we simply want to point out that the two sets 
$\{z_1,\ldots ,z_M,z_{M+1},z_{M+2}\}$ and $\{z_1,\ldots ,z_M,\xi,\eta\}$ 
are equivalent, and that we can establish a precise relation 
between the original Koba-Nielsen 
variables of the tree-level vertex 
$V_{M+2;0}$, the fixed points $\xi$ and $\eta$, and the multiplier $k$ 
of the Schottky generator at one loop. To do so, we must use the explicit 
expression of the propagator $P(x)$ and of the chosen local coordinates 
$V_i(z)$. If we choose the Lovelace 
coordinates for 
$i=M+1,M+2$, and consequently the 
propagator $P(x)$ of Ref. \cite{DFLS}, then 
a simple calculation leads to
\bea
k & = & {1-x\over{x}}~{{(z_1-z_{M+1})\,(z_M-z_{M+2})}\over{(z_1-z_{M+2})\,
(z_M-z_{M+1})}}~~~,   \nl
\xi & = & {{z_1(z_{M+1}-z_{M+2})-z_{M+1}(z_1-z_{M+2})\,k}\over 
{(z_{M+1}-z_{M+2})-(z_1-z_{M+2})\,k}}~~~, \label{changevar}  \\
\eta & = & z_{M+2}~~~.  \nonumber
\ena
Now we can exploit projective invariance to fix
\EQ
z_1 \to 1 ~~~,~~~ z_{M+1} \to k ~~~,~~~ z_{M+2} \to 0~~~,
\label{zfix}
\EN
which clearly corresponds to 
\EQ
\eta \to 0 ~~~,~~~ \xi \to \infty~~~.
\label{fixfix}
\EN
With this choice, the measure in \eq{onemeas} becomes simply
\beq
[dm]^M_1 = \prod_{i=2}^M dz_i~
dk~ {1\over k^2} \left(-{1\over {2\pi}}\ln k\right)^{-d/2}
\prod_{n=1}^\infty (1-k^n)^{2-d}~~~,
\label{onefixmeas}
\eeq
while the integration region becomes
\EQ
1\geq z_2 \geq \cdots \geq z_M \geq k \geq 0~~~.
\label{intreg}
\EN
This is the fundamental region $\Gamma_1$ over which the moduli
$z_2, \ldots , z_M$ and $k$ must be integrated, with the measure 
given in \eq{onefixmeas}.
 
We observed that the multiplier $k$ of the Schottky generator is
related to the more commonly used modular parameter ${\tilde \tau}$
by \eq{ktau}. Since $k$ is real, ${\tilde \tau}$ is purely imaginary. 
Below, we will find it more convenient to replace it with
the real variable 
\beq
\tau = -{\rm i}\pi{\tilde \tau}~~~,
\label{tau}
\eeq
which varies between 0 and $\infty$. Similarly, instead
of the Koba-Nielsen variables $z_i$, we will use the real variables
\beq
\nu_i = - \frac{1}{2} \log z_i~~~.
\label{nui}
\eeq
The integration region given by \eq{intreg} becomes, in these variables 
\beq
0 \leq \nu_2 \leq \cdots \leq \nu_{M} \leq \tau < \infty~~~. 
\label{intregnu}
\eeq
The one-loop bosonic Green
function ${\cal G}^{(1)}(z_i,z_j)$ also simplifies 
considerably using Eqs.~(\ref{zfix}) and~(\ref{fixfix}). It is given by
\beqa
{\cal G}^{(1)}(z_i,z_j) & = & 
\log (z_i - z_j) + \frac{1}{2 \log k} \left( \log \frac{z_i}{z_j} \right)^2
+ \log \left[ \prod_{n = 1}^\infty \frac{\left(1 - k^n \frac{z_j}{z_i} \right)
\left(1 - k^n \frac{z_i}{z_j} \right)}{\left(1 - k^n \right)^2} \right] \nl
& = & \log\left[- 2 \pi{\rm i}{{\theta_1\left(\frac{\rm i}{\pi}
(\nu_j-\nu_i)|\frac{\rm i}{\pi}\tau)\right)}\over{\theta'_1\left(0|
\frac{\rm i}{\pi}\tau\right)}}\right] - 
\frac{(\nu_j-\nu_i)^2}{\tau} - \nu_i - \nu_j~~~,
\label{onegreen}
\eeqa
where $\theta_1$ is the first Jacobi $\theta$ function.
It is not difficult to show that
\beq
z_i \, \partial_{z_i}{\cal G}^{(1)}(z_i,z_j) +
z_j \, \partial_{z_j}{\cal G}^{(1)}(z_i,z_j) = 1~~~.
\label{simz}
\eeq
The function ${\cal G}^{(1)}(z_i, z_j)$ is 
related to the more commonly 
used one-loop Green function $G(\nu_j - \nu_i)$ of, say, 
Ref.~\cite{GSW}, by
\beq
G(\nu_j - \nu_i) = 
{\cal G}^{(1)}(z_i(\nu_i),z_j(\nu_j)) + \nu_i + \nu_j~~~.
\label{gcalg}
\eeq
Notice that $G(\nu_j - \nu_i)$ is a function only of the difference
$\nu_j - \nu_i$, whereas ${\cal G}^{(1)}(z_i,z_j)$ 
does not have translational 
invariance in the $\nu$ plane, as is clear from \eq{simz}.
Furthermore, $G(\nu_j - \nu_i)$ is periodic on the annulus, {\it i.e.}
$G(\nu) = G(\tau + \nu)$, reminding us that the points $\nu = 0$ and 
$\nu = \tau$ represent the same physical location on the world sheet.
 
We are now in a position to write explicitly the $SU(N)$ color-ordered
planar amplitude for $M~\geq~2$ transverse gluons at one loop. It is
\bea
A^{(1)}_P (p_1,\ldots,p_M) & = & N \, {\rm Tr}(\lambda^{a_1}
\cdots \lambda^{a_M}) \, \frac{g_d^M}{(4\pi)^{d/2}} \,
(2\a')^{(M-d)/2} (-1)^{M}  \nl
& \times & \int_{\Gamma_1} \prod_{i=2}^M d\nu_i~d\tau \, 
{\rm e}^{2\tau} \, \tau^{-d/2}
\prod_{n=1}^\infty \left(1 - {\rm e}^{-2 n \tau} \right)^{2-d} \nl
& \times & \left\{\prod_{i<j} \left[ 
\sqrt{\frac{z_i\,z_j}{V'_i(0)\,V'_j(0)}}
\exp \left(G(\nu_{ji}) \right)
\right]^{2\a' p_i\cdot p_j} \right.  \label{onemaster} \\
& \times & \left. \exp \left[ \sum_{i\not=j}
\left(\sqrt{2\a'}\, p_j\cdot\ve_i 
\, \partial_i  G(\nu_{ji})+ {1\over 2} 
\ve_i\cdot\ve_j \, \partial_i \partial_j
G(\nu_{ji}) \right) \right] \right\}_{\rm m.l.}~~~,    \nonumber
\ena
where $\nu_{ji} \equiv \nu_j - \nu_i$ and $\partial_i \equiv 
\partial/\partial\nu_i$.
Notice that in the last square bracket of \eq{onemaster} we have been able
to replace everywhere ${\cal G}$ with $G$, 
since the difference is proportional
to terms that vanish because of the transversality condition 
(\ref{transverse}). If we further enforce the mass-shell condition 
(\ref{massshell}), any dependence on the
local coordinates $V_i$'s drops out. 
However, if we want to consider a formal off-shell extension of this 
amplitude, the dependence on the $V_i$'s becomes relevant. 
Our attitude is to continue off shell the gluon momenta
in an appropriate way, that we will discuss in the following sections,
essentially setting
\beq
p_i^2 = m^2~~~,
\label{offshell}
\eeq
with $m^2$ acting as an infrared regulator. Then we regard the freedom
of choosing the $V_i$'s as a gauge freedom \cite{letter}.
Clearly, the simplest situation occurs when the square root factors in 
\eq{onemaster} simplify, {\it i.e.} when 
\beq
V_i'(0) = z_i~~~.
\label{Viprime}
\eeq
With such a choice the amplitude 
in \eq{onemaster} coincides exactly with 
the one normally used~\cite{GSW}, which is entirely written
in terms of the translational invariant Green function $G(\nu)$.
The conditions (\ref{Vi}) and (\ref{Viprime}) 
are easily satisfied by choosing
for example
\beq
V_i(z) = z_i \, z + z_i~~~.
\label{gaugech}
\eeq
This is our choice for this paper.
 
We conclude this section by deriving an alternative expression for the 
Green function in \eq{gcalg}, that we will need in \secn{effact}.
 
Using Eq. (8.1.53) of Ref. \cite{GSW}, one can rewrite \eq{gcalg} as follows
\beq
G(\nu_j- \nu_i) = \log \left[ - \frac{2 \pi}{\log q} \sin \pi \left( 
\frac{\nu_j - \nu_i}{\tau} \right) \prod_{n=1}^{\infty} 
\frac{1 - 2 q^{2n} \cos 2\pi \left(\frac{\nu_j - \nu_i}{\tau} \right) + 
q^{4n}}{(1 - q^{2n})^2} \right]~~~,
\label{mtgreen}
\eeq
where $q = e^{- \pi^2/\tau}$.
By means of the identity
\beq
\log \left[ 1 + b^2 - 2 b \cos x \right] = 
- 2 \sum_{n=1}^{\infty} \frac{b^n}{n} \cos n x~~~,
\label{ident}
\eeq
it is easy to put \eq{mtgreen} in the form \cite{fratse1}
\beq
G(\nu_j- \nu_i) = 
- \sum_{n=1}^{\infty}~\frac{1 + q^{2n}}{n (1 - q^{2n})} 
\cos 2 \pi n \left( \frac{\nu_j - \nu_i}{\tau} \right) + ~\dots~~~,
\label{newmtgreen}
\eeq
where the dots stand for terms independent of $\nu_i$ and $\nu_j$, that 
will not be important in our discussion.
 
It is important to notice that the Green function in \eq{gcalg}, or in
\eq{mtgreen}, is logarithmically divergent 
in the limit $\nu_i \rightarrow
\nu_j$. This singularity is essential to get, for example, the correct pole
contributions in the various channels at tree level, as well as to extract,
in the field theory limit, the contribution of 
one-particle reducible 
diagrams at one loop, as we will see 
in Sections \ref{threepoint} and \ref{fourpoint}.
If we rewrite the Green function as in \eq{newmtgreen}, this
logarithmic singularity has not 
disappeared, but it manifests itself as a 
divergence in the sum. 
 
In what follows we will show how, in the field theory limit,
the one-loop gluon amplitudes derived from 
\eq{onemaster} develop ultraviolet 
divergences, and these divergences with our prescription reproduce exactly 
the results of the background field method in the
Feynman gauge.

\sect{The two-gluon amplitude}
\label{twopoint}
\vskip 0.5cm

Let us briefly recall the results for the two-gluon amplitude
that we already presented in Ref.~\cite{letter}. From \eq{hmaster} 
with $M=2$ we see that the color-ordered two-gluon amplitude at 
$h$ loops ($h\geq1$) is
\footnote{Notice that this expression differs from the one in Eq. (5)
of Ref.~\cite{letter} by a term proportional to 
$\ve_1\cdot p_2~\ve_2\cdot p_1$, which vanishes upon 
using momentum conservation and transversality.}
\bea
A^{(h)} (p_1,p_2) &=& N^h\,{\rm Tr}(\lambda^{a_1}\lambda^{a_2})
\,C_h\,{\cal N}_0^2 \int [dm]_h^2 \\
& \times & \left[{{\exp\left({\cal G}^{(h)}(z_1,z_2)\right)}
\over{\sqrt{V'_1(0)\,V'_2(0)}}}\right]^{2\a' p_1\cdot p_2}~
\ve_1\cdot\ve_2 \,\partial_{z_1}\partial_{z_2}
{\cal G}^{(h)}(z_1,z_2)~~~,  \nonumber
\label{twoh}
\ena
where we dropped the subscript $P$, since for $SU(N)$ the planar
diagram is the only one contributing to the two-point function.
This formula generates, in the field theory limit, 
all the loop corrections 
to the Yang-Mills two-point function. In particular, as we showed 
in Ref.~\cite{letter}, $A^{(h)}(p_1,p_2)$ can be considered as a 
master formula containing all information necessary to compute the 
multiloop Yang-Mills beta function.

Let us now specialize to one loop ($h=1$), using the explicit expressions
written in the previous section, and in particular using
$\nu_i$ as integration variables. We have
\bea
A^{(1)}(p_1,p_2) & = & N \, {\rm Tr}(\lambda^{a_1}\lambda^{a_2}) \,
\frac{g_d^2}{(4\pi)^{d/2}}(2\a')^{2-d/2}   \nl
& \times & \int_0^\infty {\cal D}\tau 
\int_0^\tau d\nu~f_2(\nu,\tau;p_1 \cdot p_2)~~~,
\label{twoone}
\ena
where we introduced the notations
\beq
{\cal D}\tau \equiv d\tau~{\rm e}^{2\tau}\,\tau^{-d/2}
\prod_{n=1}^\infty \left(1-{\rm e}^{-2n\tau} \right)^{2-d}~~~,
\label{taumeas}
\eeq
and
\beq
f_2(\nu,\tau;p_1 \cdot p_2) \equiv - {1\over 2\a'} \,
\ve_1 \cdot \ve_2 \,
{\rm e}^{2\a' p_1\cdot p_2\,G(\nu)} \, \partial_\nu^2 G(\nu)~~~.
\label{twoint} 
\eeq
For later convenience we also define the integral (see Ref.~\cite{letter})
\beq
R(s) \equiv \frac{1}{\ve_1 \cdot \ve_2} 
\frac{1}{s}  \int_0^\infty
{\cal D}\tau \int_0^\tau d\nu f_2(\nu,\tau;s)~~~.
\label{Rint}
\eeq

Notice that if the mass shell condition (\ref{massshell}) is enforced,
the two-gluon amplitude becomes ill defined, as the kinematical
prefactor vanishes, while the integral diverges.
It is then necessary to somehow continue off shell the string amplitude,
and then test that the results remain consistent in the field theory
limit. This was first attempted in Ref.~\cite{berkosrol}, where a 
prescription was found that leads to a vanishing 
wave function renormalization
in the field theory limit. On the other hand, in Ref.~\cite{letter}
we have shown that, if we formally extend the gluon momenta off shell 
according to Eqs. (\ref{onemaster}) and (\ref{offshell}), then in the 
limit $\a'\to 0$ $A^{(1)}(p_1,p_2)$ develops an 
ultraviolet divergence, 
which in turn determines a non-vanishing wave function renormalization 
constant $Z_A$, equal to the one computed in 
the background field method.

In order to see this, we follow Ref.~\cite{berkos}, and notice that 
the modular parameter $\tau$ and the coordinate $\nu$ are related to 
proper-time Schwinger parameters for the Feynman diagrams 
contributing to the two-point 
function. In particular, $t \sim \a' \tau$ and $t_1 \sim \a' \nu$, 
where $t_1$ is the proper time associated with one of the two internal 
propagators, while $t$ is the total proper time around the loop. 
In the field theory limit, with the possible exception 
of pinching 
configurations, to be discussed later,
these proper times have to remain finite, and thus the limit $\a' \to 0$
must correspond to the limit $\{ \tau, \nu \} \to \infty$ in the integrand.
This can also be checked directly by examining the integral in \eq{Rint}, 
and is most easily seen by rescaling the integration variable according to
$\hat\nu  \equiv \nu/\tau$. The field theory limit is then determined by
the asymptotic behavior of the Green function for large $\tau$, namely
\beq
G(\nu) =  - \frac{\nu^2}{\tau} + \log\left(2\sinh(\nu)\right)
- 4 \, {\rm e}^{-2\tau} \, \sinh^2(\nu) \, + \, O(e^{- 4 \tau})~~~, 
\label{lartauG}
\eeq
where $\nu$ must be taken to be finite;
in this region, we may use
\beq
G(\nu) \sim  (\hat\nu - \hat\nu^2) \tau -
\sum_{n=1}^\infty \frac{1}{n}~{\rm e}^{- 2 n\hat\nu \tau} 
- {\rm e}^{- 2 \tau (1 - \hat\nu)}
+ 2 {\rm e}^{- 2 \tau}-{\rm e}^{- 2 \tau (1 + \hat\nu)}~~~,
\label{lartaunuG}
\eeq
so that
\beq
\frac{\partial G}{\partial \nu} \sim 1 - 2 \hat\nu +
2 \sum_{n=1}^\infty {\rm e}^{- 2n \hat\nu \tau} 
- 2 {\rm e}^{- 2 \tau (1 - \hat\nu)} +
2 {\rm e}^{- 2 \tau (1 + \hat\nu)}~~~,
\label{lartaudG}
\eeq
and
\beq
\frac{\partial^2 G}{\partial \nu^2} \sim - \frac{2}{\tau} -
4 \sum_{n=1}^\infty n~{\rm e}^{- 2 n\hat\nu \tau} 
- 4 {\rm e}^{- 2 \tau (1 - \hat\nu)}
- 4 {\rm e}^{- 2 \tau (1 + \hat\nu)}~~~.
\label{lartaud2G}
\eeq
In most cases it will be sufficient to keep only the first term in
the infinite series appearing in Eqs. (\ref{lartaunuG})-(\ref{lartaud2G}). 
The only exceptions are discussed in \secn{fourpoint}.

We now substitute these results into \eq{twoone}, and keep only terms 
that remain finite when $k=e^{-2\tau}\to 0$; divergent
terms are discarded by hand, 
since they correspond to the propagation of the tachyon in the loop. 
Notice
that the divergence associated with the tachyon is not regularized by 
continuing the space-time dimension away from four, whereas all divergences
associated with gluons are turned into the usual poles of dimensional 
regularization. Notice also that by taking the large $\nu$ limit 
with $\hat\nu \sim O(1)$ we have
discarded two singular regions of integration that 
potentially contribute in
the field theory limit, namely $\nu \to 0$ and $\nu \to \tau$. 
In these regions the Green function has a logarithmic 
singularity corresponding to the
insertion of the two external states very close to 
each other, and this
singularity, as we shall see the in the next section, in general gives 
non-vanishing contributions in the field theory limit. 
However, in the case of the two-gluon amplitude, 
these regions correspond to Feynman diagrams with a loop
consisting of a single propagator, {\it i. e.} a ``tadpole''. Massless
tadpoles are defined to vanish in dimensional 
regularization, and thus we are
justified in discarding these contributions as well.

If we delete all unwanted contributions, replace the variable $\nu$ with 
${\hat \nu}\equiv\nu/\tau$, and integrate by 
parts to remove the double derivative of the Green function, 
\eq{Rint} becomes
\beq
R(s) = \int_0^\infty d\tau \int_0^1 d{\hat \nu} ~\tau^{1-d/2}\,
{\rm e}^{2\a'\,s\,({\hat\nu}-{\hat\nu}^2)\tau}
\left[(1-2{\hat\nu})^2(d-2)-8\right]~~~,
\label{limRint}
\eeq
with $s=p_1\cdot p_2$. Integration by parts is not really necessary, 
but it makes calculations easier. The integral is now elementary, and
yields
\beq
R(s) = - \Gamma\left(2-\frac{d}{2}\right)\, (- 2 \a' s)^{d/2-2}~
\frac{6-7d}{1-d}\,B\left(\frac{d}{2}-1,\frac{d}{2}-1\right)~~~,
\label{Rfin}
\eeq
where $B$ is the Euler beta function.

If we now substitute \eq{Rfin} into \eq{twoone}, we see that the
$\a'$ dependence cancels, as it must. The ultraviolet finite
string amplitude, \eq{twoone}, has been replaced by a field
theory amplitude which diverges in four space-time dimensions, 
because of the pole in the $\Gamma$ function in \eq{Rfin}.
Reverting to the customary conventions of dimensional regularization,
we set $d=4-2\e$ and $g_d=g\,\mu^\e$, 
where $g$ is dimensionless and $\mu$ an arbitrary scale.
Putting everything together we find
\bea
A^{(1)}(p_1,p_2) &=&-N\,\delta^{{a_1}{a_2}}\,\frac{g^2}{(4\pi)^2} \,
\left(\frac{4\pi\,\mu^2}{-p_1\cdot p_2}\right)^\e \,
\ve_1\cdot\ve_2\,p_1\cdot p_2 \nl
& \times &\Gamma(\e)\,\frac{11-7\e}{3-2\e}\,
B(1-\e,1-\e)~~~.
\label{twofin}
\ena
\eq{twofin}) is exactly equal to the gluon vacuum
polarization of Yang-Mills theory, computed
with the background field method, in Feynman gauge, with dimensional 
regularization.

{}From \eq{twofin} we see that the 
ultraviolet divergence at $d=4$ can be removed
by a wave function renormalization, with a minimal subtraction 
$Z$ factor given by
\beq
Z_A= 1 + N\,\frac{g^2}{(4\pi)^2}\,\frac{11}{3}\,\frac{1}{\e}~~~.
\label{za}
\eeq
While this result is what we expected, to make sure that our prescription
is consistent we need to go on to compute the three and four-point
renormalizations as well, and verify that 
gauge invariance is preserved.
This we will do in the next two sections.

\sect{The three-gluon amplitude}
\label{threepoint}
\vskip 0.5cm

The color-ordered amplitude for three transverse gluons at tree
level is given by \eq{treemaster} with $M=3$, that is
\bea
A^{(0)}(p_1,p_2,p_3) & = & - \, 4 \, g_d \, 
{\rm Tr}(\lambda^{a_1}\lambda^{a_2}\lambda^{a_3})~\Big(
\ve_1\cdot\ve_2\,p_2\cdot\ve_3 \nonumber \\
& & + \, \ve_2\cdot\ve_3\,p_3\cdot\ve_1
+ \ve_3\cdot\ve_1\,p_1\cdot\ve_2 +
O(\a') \Big)~~~,
\label{threetree}
\ena
where  $O(\a')$ stands for string corrections,
proportional to $\a'$, such as
$\ve_1\cdot p_2 \, \ve_2\cdot p_3 \, \ve_3\cdot p_1$.
Since we are ultimately interested in the field theory limit,
it is not necessary to write such terms explicitly. 
As is well known, when $\a' \to 0$, $A^{(0)}(p_1,p_2,p_3)$
in \eq{threetree} becomes the standard Yang-Mills color-ordered 
three-gluon amplitude, and the reader may easily verify
the correctness of the normalization, taking into account 
our conventions.

The one-loop correction to \eq{threetree} is given by \footnote{Once
again we omit the subscript $P$, as only the 
planar one-loop diagram
contributes for $SU(N)$.}
\bea
A^{(1)}(p_1,p_2,p_3) & = & - \, N \, 
{\rm Tr}(\lambda^{a_1}\lambda^{a_2}\lambda^{a_3}) \,
\frac{g_d^3}{(4\pi)^{d/2}} \, (2\a')^{2-d/2} \nl
& \times & \!\!\! \int_0^\infty {\cal D}\tau \int_0^\tau \!d\nu_3 
\int_0^{\nu_3}\!d\nu_2 ~f_3(\nu_2,\nu_3,\tau)~~~,
\label{threeone}
\ena
where
\bea
f_3(\nu_2,\nu_3,\tau) & \equiv &
{\rm e}^{2\a'p_1\cdot p_2\,G(\nu_{2})}~
{\rm e}^{2\a'p_2\cdot p_3\,G(\nu_{32})}~
{\rm e}^{2\a'p_3\cdot p_1\,G(\nu_3)} \nl
& \times & \Bigg\{ \left[
- \, \ve_1\cdot\ve_2 \, \partial_2^2 G(\nu_2) \left(
p_1\cdot\ve_3 \, \partial_3 G(\nu_3) + 
p_2\cdot\ve_3 \, \partial_3 G(\nu_{32}) \right) \right.  \nl
& & + \left. \, \ve_2\cdot\ve_3 \, \partial_3^2 G(\nu_{32}) 
\left(p_2\cdot\ve_1 \, \partial_2 G(\nu_2) +
p_3\cdot\ve_1 \, \partial_3 G(\nu_3) \right) \right.  \nl
& & + \left. \, \ve_1\cdot\ve_3 \, \partial_3^2 G(\nu_3) \left(
p_3\cdot\ve_2 \, \partial_3G(\nu_{32}) -
p_1\cdot\ve_2 \, \partial_2G(\nu_2) \right) \right]  \nl
& & + \, \, O(\a') \Bigg\}~~~.
\label{threeint}
\ena
Our task is now to compare $A^{(1)}(p_1,p_2,p_3)$ with 
$A^{(0)}(p_1,p_2,p_3)$ in the limit $\a'\to 0$, 
and thus derive the three-gluon vertex renormalization constant 
at one loop.

\begin{figure}
\begin{center}
\begin{picture}(10000,8000)
\drawloop\gluon[\N 8](0,1100)
\drawline\gluon[\W\FLAT](50,1500)[3]
\global\advance \gluonbackx by -1100
\put(\gluonbackx,\gluonbacky){1}
\drawline\gluon[\NE\REG](4000,3000)[2]
\global\advance \gluonbackx by 750
\put(\gluonbackx,\gluonbacky){2}
\drawline\gluon[\SE\REG](4000,0)[2]
\global\advance \gluonbackx by 750
\put(\gluonbackx,\gluonbacky){3}
\end{picture}
\vskip 1cm
\caption{\label{I123} Representative diagram for the type $I$ region 
of the three-gluon amplitude.}
\end{center}
\end{figure}
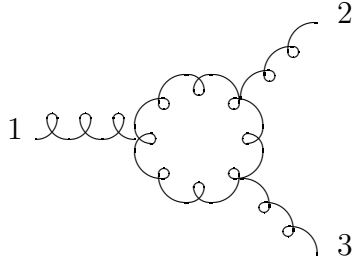

\begin{figure}
\begin{center}
\begin{picture}(10000,8000)
\drawloop\gluon[\N 8](0,1100)
\drawline\gluon[\W\FLAT](50,1500)[3]
\global\advance \gluonbackx by -1100
\put(\gluonbackx,\gluonbacky){1}
\drawline\gluon[\NE\REG](4800,1500)[3]
\global\advance \gluonbackx by 750
\put(\gluonbackx,\gluonbacky){2}
\drawline\gluon[\SE\REG](4800,1500)[3]
\global\advance \gluonbackx by 750
\put(\gluonbackx,\gluonbacky){3}
\end{picture}
\vskip 1cm
\caption{\label{II23} Representative diagram for the type $II$ region 
$\hat\nu_{32}=O(\tau^{-1})$ of the three-gluon amplitude.}
\end{center}
\end{figure}
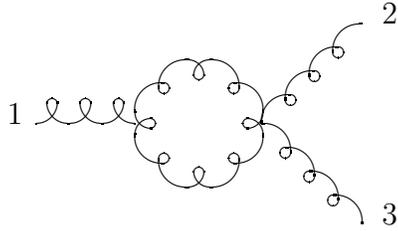

\begin{figure}
\begin{center}
\begin{picture}(10000,6000)
\drawloop\gluon[\N 8](0,1100)
\drawline\gluon[\W\FLAT](50,1500)[3]
\global\advance \gluonbackx by -1100
\put(\gluonbackx,\gluonbacky){1}
\drawline\gluon[\E\FLIPPEDFLAT](4800,1500)[2]
\drawline\gluon[\NE\REG](\gluonbackx,\gluonbacky)[3]
\global\advance \gluonbackx by 750
\put(\gluonbackx,\gluonbacky){2}
\drawline\gluon[\SE\REG](\gluonfrontx,\gluonfronty)[3]
\global\advance \gluonbackx by 750
\put(\gluonbackx,\gluonbacky){3}
\end{picture}
\vskip 1cm
\caption{\label{III23} Representative diagram for the pinching region 
$\hat\nu_{32}=O(\tau^{-2})$ of the three-gluon amplitude.}
\end{center}
\end{figure}
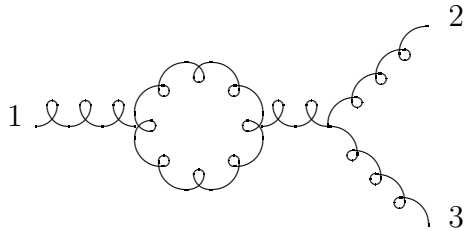

A careful analysis of \eq{threeone} allows us to distinguish
three different types of regions in the 
integration domain that contribute to the field theory limit,
and to associate the three corresponding contributions 
to the three classes of Feynman diagrams 
depicted in Figs. \ref{I123}-\ref{III23}.
To understand this, let us concentrate on two of the punctures, 
say $\nu_2$ and $\nu_3$, and follow the reasoning of 
\secn{twopoint}, which again
leads us to identify the field theory limit of the amplitude
(\ref{threeone}) with the limit
$\tau \to \infty$ in the integrand (\ref{threeint}). 
There are now, however, three 
different ways of taking this limit
that give finite contributions. First of all, one can let
\beq
\tau \rightarrow \infty ~~~,~~~~~~
\hat\nu_3 - \hat\nu_2 \equiv \hat\nu_{32} = O(1)~~~,
\label{reg1}
\eeq
which corresponds to shrinking the width of the annulus, 
while keeping the punctures widely separated. 
This region is strictly analogous to the only region that contributes 
to the two-point function,
and can be discussed along the same lines of \secn{twopoint}. 
We refer to it as a region of type $I$. As we shall see, 
this region will generate the contribution of 
one-particle irreducible diagrams with three propagators 
in the loop, and thus only three-point vertices, as shown in \fig{I123}. 

The second relevant region is
\beq
\tau \rightarrow \infty ~~~,~~~~~~
\hat\nu_{32} = O(\tau^{-1})~~~.
\label{reg2}
\eeq
Here the $\nu$ coordinates are still widely separated, since
$\nu_3 - \nu_2 = O(1)$, but the proper time of the 
corresponding propagator
vanishes as $\tau \to \infty$, so that the two punctures
come together in the field theory limit. 
This region, which we call of type $II$, will generate the
contribution of a diagram with a four-point vertex 
and only two propagators 
in the loop, as shown in \fig{II23}. Notice that since the
$\nu$'s are still widely separated, the expansion given in \eq{lartauG} 
for the Green function $G(\nu_{32})$ 
can still be applied, as we are still away from 
the singularity corresponding to $\nu_2 = \nu_3$.

Finally, the third relevant region is the pinching region, defined by
\beq
\tau \rightarrow \infty ~~~,~~~~~~
\hat\nu_{32} = O(\tau^{-2})~~~,
\label{reg3}
\eeq
so that in the limit $\tau \to \infty$ one approaches 
the logarithmic singularity of 
$G(\nu_{32})$ as $\nu_{32} \to 0$. 
In this region, which we call of type $III$, the expansion 
in \eq{lartauG} is no longer applicable, and one must use instead
\beq
G(\nu) \sim \log(2 \, \nu)~~~,
\label{pinchG}
\eeq
for the Green function and its derivatives. As shown already
in Ref.~\cite{berkos}, pinching two variables is equivalent 
to the introduction of a propagator attached to the loop, and 
thus generates an exchange diagram in the 
channel determined by the 
two pinched punctures. When such an exchange involves only gluons, the
corresponding contribution to the amplitude in the limit
$\a' \to 0$ certainly corresponds to a 
one-particle reducible diagram, as shown in
\fig{III23}. However, there may be cases in which after 
pinching, the resulting exchange involves also the tachyon of the 
bosonic string. This leads to a divergence which we
discard by hand, as we did for the divergence due to tachyons circulating 
in the loop. As we shall see in \secn{fourpoint}, there are circumstances
in which the propagation of a tachyon in a pinched configuration can give
a contribution that survives in the field theory limit, in the form of a 
contact interaction among four gluons. Such contributions are properly
dealt with by assigning them to a type-$II$ region, together with all
other contributions corresponding to four-point vertices directly
attached to the loop.

Since we are interested in minimal subtraction renormalization 
constants, in the following we will
concentrate on the contributions that diverge as 
$d \to 4$. They are easy to isolate, as they can be produced only by an
integral over the modular parameter $\tau$ containing the factor
$\tau^{1 - d/2}$, that gives $\Gamma(2 - d/2)$ 
(lower powers of $\tau$ are
suppressed by corresponding extra powers of $\a'$). 
Since each $\nu$ integration generates a factor of $\tau$
through the change of variable $\nu\rightarrow\hat\nu$,
we actually need the square bracket in \eq{threeint} to generate
a single negative power of $\tau$, in order to produce a 
divergent integral. 

In order to display the relevant contributions, let's focus our 
attention on the term of \eq{threeone}
proportional to $\ve_1\cdot\ve_2$ which, up to the
constant prefactors, is given by
\bea
J(p_1, p_2, p_3) & = & \int_0^\infty d\tau~{\rm e}^{2\tau} \,
\tau^{-d/2} \prod_{n=1}^\infty
\left(1-{\rm e}^{-2n\tau}\right)^{2-d}
\int_0^\tau d\nu_3 \int_0^{\nu_3} d\nu_2 \nl
& \times & {\rm e}^{2\a'p_1\cdot p_2 \, G(\nu_{2})}~
{\rm e}^{2\a'p_2\cdot p_3 \, G(\nu_{32})}~
{\rm e}^{2\a'p_3\cdot p_1 \, G(\nu_3)}  \label{focus}  \\
& \times & \Bigg[ - \ve_1\cdot\ve_2 \, 
\partial_2^2 G(\nu_2) \Big(
p_1\cdot\ve_3 \, \partial_3 G(\nu_3) +
p_2\cdot\ve_3 \, \partial_3 G(\nu_{32}) 
\Big) \Bigg]~~~.  \nonumber
\ena
The other terms of $A^{(1)}(p_1, p_2, p_3)$ 
can be obtained by cyclic
permutations from $J(p_1, p_2, p_3)$. 

We start from the region of type $I$ defined by
\beq
\tau\to\infty\;\;,~~~~~~\hat\nu_{i+1,i}=O(1)~~~
{\rm for~all}~~1\leq i\leq3~~~,
\label{reg1tau}
\eeq
where $\hat\nu_{43}\equiv 1-\hat\nu_3$.
Here we can use the first term of the expansion of 
$\partial_\nu^2 G(\nu)$, \eq{lartaud2G}, in order 
to have the right power of $\tau$. Then we notice that
no term in the expansion of $\partial_\nu G(\nu)$,
\eq{lartaudG}, can exactly compensate the factor of $k^{-1} = 
{\rm e}^{2 \tau}$ in the measure, that signals the presence of the 
tachyon in the loop. 
Thus we are forced to pick up a factor of $k$ from the 
expansion of $\prod_n (1-k^{n})^{2-d}$. Singling out these terms in
the expansion of \eq{focus}, we find that the total contribution to 
the ultraviolet divergence from region $I$ can be written as
\bea 
J(p_1, p_2, p_3)\Big|_I & = & - 2 \, (d - 2) \, \ve_1\cdot\ve_2 
\, p_2\cdot\ve_3 \int_0^\infty d\tau~ \tau^{2 - d/2} 
\left(-{2\over\tau}\right) \int_0^1 d \hat\nu_3 \int_0^{\hat\nu_3} 
d \hat\nu_2 \nl 
& \times & (\hat\nu_3 - \hat\nu_{32})~
{\rm e}^{2\a'p_1\cdot p_2\,G(\nu_2)}~
{\rm e}^{2\a'p_2\cdot p_3\,G(\nu_{32})}~
{\rm e}^{2\a'p_3\cdot p_1\,G(\nu_3)}~~~.
\label{focus1}
\ena
If we are only interested in the coefficient of the $1/\e$ pole,
we can simply replace the exponentials with an infrared cutoff of
the form $\exp (- 2 \a' m^2 \tau)$. Then
\beq 
J(p_1, p_2, p_3)\Big|_I = {4\over3} \, \ve_1\cdot\ve_2
\, p_2\cdot\ve_3 \, {1\over\e} + O(\e^0)~~~.
\label{onediv}
\eeq
We can now check that the region of moduli space just considered
in fact corresponds to the diagrams in field theory with three
propagators in the loop. Employing the usual Feynman rules of the 
background field method~\cite{abbott} one easily finds that these
diagrams give in fact the same result as \eq{onediv}.

Let us now turn our attention to the regions of type $II$. 
There are three such regions, all characterized by $\tau\to\infty$, while
\beq
{\hat\nu}_{32} = O(\tau^{-1})~~~,
\label{reg2tau}
\eeq
or
\beq
1-{\hat\nu}_{3} = O(\tau^{-1})~~~,
\label{reg2tau1}
\eeq
or
\beq
{\hat\nu}_{2} = O(\tau^{-1})~~~.
\label{reg2tau2}
\eeq

\begin{figure}
\begin{center}
\begin{picture}(7000,6000)
\drawloop\gluon[\N 8](0,1100)
\drawline\gluon[\NE \REG](4000,3000)[2]
\global\advance \gluonbackx by 750
\put(\gluonbackx,\gluonbacky){2}
\drawline\gluon[\W \FLAT](200,400)[3]
\global\advance \gluonbackx by -1100
\put(\gluonbackx,\gluonbacky){1}
\drawline\gluon[\S \REG](\gluonfrontx,\gluonfronty)[3]
\global\advance \gluonbackx by 750
\put(\gluonbackx,\gluonbacky){3}
\end{picture}
\vskip 1cm
\caption{\label{II31} Representative diagram for the type $II$ region 
$1-\hat\nu_{3}=O(\tau^{-1})$ of the three-gluon amplitude.}
\end{center}
\end{figure}
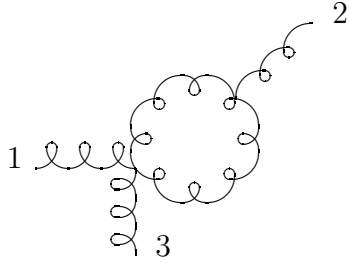

Region (\ref{reg2tau}) corresponds to a
vanishing proper time between the punctures $\nu_2$ and 
$\nu_3$ and can 
be associated with the diagram in \fig{II23}. Similarly, if we 
keep in mind that the two points $\nu_1 = 0$ and $\tau$ are to be 
identified, we realize that the region 
in \eq{reg2tau1} corresponds to a vanishing proper
time between $\nu_3$ and $\nu_1$, and thus it can be 
associated with the diagram represented in 
\fig{II31}. Finally, the third region, \eq{reg2tau2},
corresponds to a vanishing proper time between $\nu_1$ and $\nu_2$
and can be associated with a diagram with two propagators and a
four-point vertex between the gluons 1 and 2.

In order to see how these regions contribute to $J(p_1,p_2,p_3)$, 
we use for $\partial_\nu G(\nu)$ and $\partial_\nu^2 G(\nu)$ the terms
of their expansions, Eqs. (\ref{lartaudG}) and (\ref{lartaud2G}),
that are of the form $k^{\hat\nu}$.
They combine to give 
\beq
- 8\, \ve_1\cdot\ve_2  \,p_1\cdot\ve_3 
\, \left( k^{(1 - \hat\nu_{32})} - 
k^{(1 + \hat\nu_{32})}\right) + \ldots 
\label{ktonu}
\eeq
in region (\ref{reg2tau}), and 
\beq
 8\, \ve_1\cdot\ve_2 \, p_2\cdot\ve_3 
\, \left( k^{ \hat\nu_3} - 
k^{(2-\hat\nu_3 )} \right) +\ldots
\label{ktonu1}
\eeq
in region (\ref{reg2tau1}). 
For general values of $\hat\nu_2$ and $\hat\nu_3$ such terms, as
well as all the others which we have not exhibited, fail 
to compensate the leading factor of $k^{-1}$ from the measure.
However, precisely in the regions (\ref{reg2tau}) and (\ref{reg2tau1}) 
the terms written in Eqs. (\ref{ktonu})
and (\ref{ktonu1}) become proportional to $k$, and thus give a 
finite contribution. Once again, if we are only interested in 
the ultraviolet
divergence we can replace the exponential of the Green functions by
an infrared cutoff as above, whereupon the integral 
over the punctures
becomes trivial. Indeed, the expression in \eq{ktonu} yields a
factor of $(8/\tau) (\ve_1
\cdot\ve_2 \,p_1\cdot\ve_3)$
from region (\ref{reg2tau}), while \eq{ktonu1}
produces a factor of $(-8/\tau) (\ve_1
\cdot\ve_2 \,p_2\cdot\ve_3)$ from region (\ref{reg2tau1}).
On the other hand, one can check that 
region (\ref{reg2tau2}) does not give any
contribution to the ultraviolet divergence of
$J(p_1, p_2, p_3)$. In fact, in this case it is not
possible to produce the right power of the modular parameter
$\tau$ from the $\hat\nu$ integration.
  
Recalling that $p_1\cdot\ve_3=-p_2\cdot\ve_3$, the total 
contribution to $J(p_1, p_2, p_3)$ from regions of type 
$II$ is then
\beq 
J(p_1, p_2, p_3)\Big|_{II} = - 16 \,\ve_1\cdot\ve_2 \,p_2\cdot\ve_3 \,
{1\over\e} + O(\e^0)~~~,
\label{twodiv}
\eeq
and one can verify that this is exactly the 
contribution from
diagrams with a four-gluon vertex in the background field method,
as conjectured.

Finally, we turn to the analysis of the pinching regions.
In our case, there are three such regions, namely $\tau\to\infty$ with
\beq
{\hat\nu}_{2} = O(\tau^{-2})~~~,
\label{nu2to1}
\eeq
or
\beq
{\hat\nu}_{32} = O(\tau^{-2})~~~,
\label{nu2to3}
\eeq
or
\beq
1-{\hat\nu}_{3} = O(\tau^{-2})~~~,
\label{nu3to1}
\eeq
as dictated
by cyclic symmetry and periodicity on the annulus. One can easily
see that in all those regions
at least one of the Green functions in \eq{focus} is singular.

Let us consider, for example, the first pinching, {\it i.e.}
$\nu_2 \to 0$. 
Since $\nu_2$ is localized in a neighbourhood
of $0$, we can replace the integral $\int_0^{\nu_3} d \nu_2$ with
an integral $\int_0^\eta d \nu_2$, where $\eta$ is an arbitrary small
number. Further, we can use the approximation in \eq{pinchG} for the
Green function $G(\nu_2)$ and its derivatives.
After this is done, we can expand
$G(\nu_{32})$ in powers of $\nu_2$, which turns the amplitude 
$J(p_1,p_2,p_3)$ into an infinite series. 
The $n$-th term of this series is proportional to an integral of the form
\beq
C_n \equiv \int_0^\eta d\nu_2~\nu_2^{n-2+2\a'p_1\cdot p_2}
\label{pinchser}
\eeq
with $n \geq 0$. After a suitable analytic continuation in the momenta 
to insure convergence, we get
\beq
C_n = \frac{\eta^{n-1+2\a'p_1\cdot p_2}}{n-1+2\a'p_1\cdot p_2}~~~.
\label{pinchpol}
\eeq
Therefore, when the pinching $\nu_2\to 0$ is performed, the
amplitude $J(p_1,p_2,p_3)$ becomes an infinite sum over
all possible string states that are exchanged in the $(12)$-channel, 
$n=0$ corresponding to the tachyon, $n=1$ to the gluon and so on.
In the case of the three-gluon amplitude,
the exchange of a tachyon does not give any contribution: in fact,
for $\nu_2\to 0$
\bea 
&&
\partial_2^2 G(\nu_2) \Big( p_1\cdot\ve_3\,\partial_3 G(\nu_3)
+ p_2\cdot\ve_3\,\partial_3 G(\nu_{32})\Big) \label{notach} \\ 
&&~~~~~~= -{1\over\nu_{2}^2}~
(p_1 + p_2)\cdot\ve_3 \,\partial_3 G(\nu_{3}) + 
{1\over \nu_{2}}~ p_2\cdot\ve_3 \,\partial_3^2 G(\nu_{3})
+\ldots~~~.
\nonumber 
\ena
Using the trasversality of the external states, we see that
the coefficient of the quadratic divergence ${1\over\nu_2^2}$ is
zero. Even if this tachyon contribution had not been vanishing, 
we would have discarded it, according to our prescription.
On the other hand, the gluon term survives in the field theory
limit, and actually contributes to the ultraviolet 
divergence, as expected: in fact, the single
pole in $\nu_2$ generates, through the change of variables to
$\hat\nu_2$, the negative power of $\tau$ needed 
for the integral to diverge. 
All other terms in the series, corresponding to $n \geq 2$, 
and to states 
whose mass becomes infinite as $\a' \to 0$, vanish in the field theory 
limit. 

\begin{figure}
\begin{center}
\begin{picture}(7000,9000)
\drawloop\gluon[\N 8](0,1100)
\drawline\gluon[\SE \REG](4000,0)[2]
\global\advance \gluonbackx by 750
\put(\gluonbackx,\gluonbacky){3}
\drawline\gluon[\NW \REG](1010,3000)[2]
\drawline\gluon[\W \FLAT](\gluonbackx,\gluonbacky)[3]
\global\advance \gluonbackx by -1100
\put(\gluonbackx,\gluonbacky){1}
\drawline\gluon[\N \REG](\gluonfrontx,\gluonfronty)[3]
\global\advance \gluonbackx by -1100
\put(\gluonbackx,\gluonbacky){2}
\end{picture}
\vskip 1.5cm
\caption{\label{III12} Representative diagram for the pinching region 
$\nu_{2}\to 0$ of the three-gluon amplitude.}
\end{center}
\end{figure}

Keeping this in mind, and collecting all relevant factors, we find that 
the contribution to $J(p_1,p_2,p_3)$ from the pinching $\nu_2\to 0$ is
\beq
J(p_1,p_2,p_3)\Big|_{\nu_2\to 0}  = 
\frac{(p_1+p_2)\cdot p_3}{p_1\cdot p_2}~
R\left[(p_1+p_2)\cdot p_3\right] \,
\ve_1\cdot\ve_2 \, p_2\cdot\ve_3~~~, \label{pinch}
\eeq
where $R$ is the integral defined in \eq{Rint}. Notice that
\eq{pinch} contains a ratio of momentum invariants which are vanishing
on shell. The appearance of such ratios in string amplitudes, 
in the
corners of moduli space corresponding to loops isolated on external 
legs, is
a well-known fact, which for example motivated the work of 
Ref.~\cite{berkosrol}. As we already remarked in Ref.~\cite{letter}, 
this ``$0/0$'' ambiguity is similar to the one that appears in 
the unrenormalized connected
Green functions of a massless field theory, if the external legs
are kept on the mass-shell and divergences are 
regularized with dimensional
regularization. There, such ambiguity is removed by 
going off shell; similarly
here, we formally continue the gluon 
momenta off the mass shell, following 
the same prescription we already adopted in the previous section for the 
two-gluon amplitude, {\it i.e.} we put off shell the  
momentum of the gluon attached to the loop, according to
\beq 
p_3^2 = (p_1 + p_2)^2 = m^2~~~.
\label{off3}
\eeq
In the present case, however, this prescription is not enough 
to remove the ambiguity in \eq{pinch}, and we must further 
decide whether and how to continue off shell 
also the other 
gluon momenta, $p_1$ and $p_2$. Here we rely on the assumption,
substantiated by the results obtained so far, that string amplitudes lead
to field theory amplitudes calculated with the background field method.
As was shown in Ref.~\cite{abbgrisch}, 
to calculate {\it amplitudes} with the 
background field method it is necessary to 
treat one-particle irreducible and one-particle reducible diagrams 
quite differently. $S$-matrix elements
are obtained by first calculating one-particle irreducible vertices to the
desired order, and then gluing them together with propagators that can only
be defined when the gauge for the {\it background field} has been fixed.
As noticed also in Ref.~\cite{berdun}, the gauge chosen for the 
background field is quite independent from the gauge that 
had been chosen for the quantum field, in our case the Feynman gauge. 
This leads us to interpret \eq{pinch}
as a one-loop, one-particle irreducible two-point function, whose
momentum must be continued off shell according to \eq{off3}, glued to a 
tree-level three-point vertex, for which no off-shell continuation is
necessary. We thus keep $p_1^2 = p_2^2 = 0$, which, using momentum 
conservation, implies
\beq
\frac{(p_1+p_2)\cdot p_3}{p_1\cdot p_2}=-2~~~.
\label{off12}
\eeq
Then, with this prescription, we find
\bea
J(p_1,p_2,p_3)\Big|_{\nu_2\to 0}  &=& -2\,
R(-m^2)\,
\ve_1\cdot\ve_2 \, p_2\cdot\ve_3\nonumber \\
&=& \frac{44}{3}\,\ve_1\cdot\ve_2\,p_2\cdot\ve_3\,\frac{1}{\epsilon} +
O(\epsilon^0)~~~,
\label{pinchfin}
\ena
where we have used \eq{Rfin} with $d=4-2\epsilon$.
One can easily check that this is exactly the contribution of
one-particle reducible diagrams with the loop isolated on the third leg
as shown in \fig{III12}.

The other two pinchings, \eq{nu2to3} and \eq{nu3to1},
can be analyzed in a similar way with obvious changes of labels. 
One finds that they lead to the same
result of \eq{pinchfin} and can be associated with the
diagrams represented in \fig{III23} and \fig{III31}
with the loop isolated on the first and the second leg 
respectively.
The total contribution to $J(p_1,p_2,p_3)$ from the type $III$ region 
is then
\beq
J(p_1,p_2,p_3)\Big|_{III} = 
44\, \ve_1\cdot\ve_2\,p_2\cdot\ve_3\,\frac{1}{\epsilon} +
O(\epsilon^0)~~~.
\label{pinchtot}
\eeq

As we have mentioned before, the other two terms
of the amplitude (\ref{threeone}) that are proportional
to $\ve_2\cdot\ve_3$ and $\ve_3\cdot\ve_1$, can be simply obtained from
$J(p_1,p_2,p_3)$ by cyclic permutations of indices.
We can now write the field theory limit of
the full one-loop amplitude $A^{(1)}(p_1,p_2,p_3)$ and 
extract from it the vertex renormalization
constant. 
Reinstating all normalization factors, and  summing the
contributions from regions $I$, $II$ and $III$, we find that the
total divergence is  
\beq 
A^{(1)}(p_1,p_2,p_3)\Big|_{\rm div} = 
2N\,\frac{g^2}{(4\pi)^2}\,\frac{11}{3}\,\frac{1}{\e}
\, A^{(0)}(p_1,p_2,p_3)~~~.
\label{totdiv3}
\eeq
Comparing with what we expect from field theory for a
connected three-point amplitude, namely
\beq
A^{(1)}(p_1,p_2,p_3)\Big|_{\rm div} = (Z_3^{-1}\,Z_A^3 - 1) \,
A^{(0)}(p_1,p_2,p_3)~~~,
\label{expect}
\eeq
and using the result of \eq{za},
we are led to 
\beq 
Z_3= 1 + N\,\frac{g^2}{(4\pi)^2}\,\frac{11}{3}\,\frac{1}{\e}=Z_A~~~.
\label{z3=za}
\eeq
The Ward identity 
\beq 
Z_3 = Z_A~~~
\label{ward1} 
\eeq 
is typical of the background field method. Thus our present result
is a confirmation of what we have already found in 
\secn{twopoint}, and provides also a non-trivial consistency check on 
the whole procedure. 

\begin{figure}
\begin{center}
\begin{picture}(7000,7000)
\drawloop\gluon[\N 8](0,1100)
\drawline\gluon[\NE \REG](4000,3000)[2]
\global\advance \gluonbackx by 750
\put(\gluonbackx,\gluonbacky){2}
\drawline\gluon[\SW \REG](1010,0)[2]
\drawline\gluon[\W \FLAT](\gluonbackx,\gluonbacky)[3]
\global\advance \gluonbackx by -1100
\put(\gluonbackx,\gluonbacky){1}
\drawline\gluon[\S \REG](\gluonfrontx,\gluonfronty)[3]
\global\advance \gluonbackx by 750
\put(\gluonbackx,\gluonbacky){3}
\end{picture}
\vskip 3cm
\caption{\label{III31} Representative diagram for the pinching region 
$\nu_{3}\to\tau$ of the three-gluon amplitude.}
\end{center}
\end{figure}

Further evidence of this fact is given by the following
observation: since the contributions 
from regions $I$ and $II$ are represented 
by the one-particle irreducible diagrams of Figs. \ref{I123}, \ref{II23} and 
\ref{II31}, while those
of the pinching regions are given by the reducible diagrams of 
Figs. \ref{III23}, \ref{III12} and \ref{III31}, 
we are led to believe that the sum of the contributions from
regions $I$ and $II$ should be the proper three-point vertex
and give directly 
the renormalization constant $Z_3$. This is exactly what happens.
In fact, 
\bea
A^{(1)}(p_1,p_2,p_3)\Big|_{\rm 1PI} &\equiv&
A^{(1)}(p_1,p_2,p_3)\Big|_{I} + A^{(1)}(p_1,p_2,p_3)\Big|_{II} 
\nonumber \\
&=&
- N\,\frac{g^2}{(4\pi)^2}\,\frac{11}{3}\,\frac{1}{\e}
\, A^{(0)}(p_1,p_2,p_3)
+ O(\e^0)~~~,
\label{1pi3}
\ena
which once again leads to the background field Ward identity
(\ref{z3=za}).

In order to obtain these results, we could have 
adopted a slightly different 
procedure, integrating by parts the double derivatives 
of the Green functions
as we did in \secn{twopoint}. Doing that one arrives at 
the same result
found by Bern, Kosower and Roland in Ref.~\cite{berkosrol} (they used
a different string model, but that does not have any influence
on the field theory limit). Integration by parts moves all 
contributions to the renormalization constants 
to the pinching regions, and
thus spoils the identification of the different 
regions of moduli space with
classes of Feynman diagrams, which is clearly displayed here. However,
if one uses our prescription, \eq{off3} and \eq{off12},
to remove the ambiguity associated with the pinching, \eq{z3=za} is 
again obtained.

As a final consistency check, we will verify the gauge invariance
of our prescription, by calculating in the following section
the renormalization of the four-point
vertex, and showing that it is properly related to the renormalization
of the gauge coupling.

\sect{The four-gluon amplitude}
\label{fourpoint}
\vskip 0.5cm

In order to obtain the renormalization constant $Z_4$ of the four-point
vertex, we need to compare the four-gluon amplitude at one loop with
the tree-level expression,
\bea 
A^{(0)}(p_1,p_2,p_3,p_4) & = & 4~g_d^2~{\rm Tr}
(\lambda^{a_1}\lambda^{a_2}\lambda^{a_3}\lambda^{a_4})~\Bigg(
\ve_1\cdot\ve_2\,\ve_3\cdot\ve_4 \, {p_1\cdot p_3\over p_1\cdot p_2}
\nonumber \\
& + & \ve_1\cdot\ve_3\,\ve_2\cdot\ve_4
+ \ve_1\cdot\ve_4\,\ve_2\cdot\ve_3 \, {p_1\cdot
p_3\over p_2\cdot p_3} \, + \, \ldots \, \Bigg)~~~.
\label{fourtree}
\ena
Here terms of the form $(\ve\cdot\ve){(\ve
\cdot p)}^2$, as well as higher orders in $\a'$, have not been written
explicitly, since they will not play any role in our discussion. 
A more complete expression for $A^{(0)}(p_1,p_2,p_3,p_4)$
can be found in \eq{afour} of Appendix A.

The color-ordered planar amplitude with four gluons at one loop is given 
by \eq{onemaster} in the case $M=4$. Notice that in the open bosonic
string the complete one-loop four-gluon amplitude 
receives a contribution 
also from the non-planar diagram, where two gluons are emitted at one
boundary of the annulus and two at the other. However, the non-planar
diagram does not contribute to the renormalization constant $Z_4$,
so we need not consider it here.

To simplify the analysis, we will again focus on a single term of 
the full four-gluon amplitude. 
The simplest choice is the term proportional 
to $\ve_1\cdot\ve_3\,\ve_2\cdot\ve_4$: in fact, 
for the chosen color ordering, at tree-level this term is generated
only by the 1PI diagram given by the four-point vertex
\footnote{At the end of this section we will briefly
comment on the term proportional to 
$\ve_1\cdot\ve_2\,\ve_3\cdot\ve_4$,
which does not share this property.}.
Thus, we will calculate the contributions 
from the three regions described in the previous section
to the integral
\bea
J(p_1, p_2, p_3,p_4) & = & \int_0^\infty d\tau~{\rm e}^{2\tau} \,
\tau^{-d/2} \prod_{n=1}^\infty
\left(1-{\rm e}^{-2n\tau}\right)^{2-d}\int_0^\tau d  \nu_4
\int_0^{\nu_4} d\nu_3 \int_0^{\nu_3} d\nu_2  \nl
& \times & {\rm e}^{2\a'p_1\cdot p_2 \, G(\nu_{2})}~
{\rm e}^{2\a'p_1\cdot p_3 \, G(\nu_{3})}~
{\rm e}^{2\a'p_1\cdot p_4 \, G(\nu_{4})}~\label{focus4}\\
& \times &
{\rm e}^{2\a'p_2\cdot p_3 \, G(\nu_{32})}~ 
{\rm e}^{2\a'p_2\cdot p_4 \, G(\nu_{42})}~
{\rm e}^{2\a'p_3\cdot p_4 \, G(\nu_{43})} \nl
& \times &\ve_1\cdot\ve_3 \,\ve_2\cdot\ve_4 \,
\partial_3^2 G(\nu_3)
\partial_4^2 G(\nu_{42})~~~.
\nonumber
\ena

\begin{figure}
\begin{center}
\begin{picture}(7000,8000)
\drawloop\gluon[\N 8](0,1100)
\drawline\gluon[\SW\REG](1020,0)[2]
\global\advance \gluonbackx by -1100
\put(\gluonbackx,\gluonbacky){1}
\drawline\gluon[\NW\REG](1020,3000)[2]
\global\advance \gluonbackx by -1100
\put(\gluonbackx,\gluonbacky){2}
\drawline\gluon[\NE\REG](4000,3000)[2]
\global\advance \gluonbackx by 1100
\put(\gluonbackx,\gluonbacky){3}
\drawline\gluon[\SE\REG](4000,0)[2]
\global\advance \gluonbackx by 1100
\put(\gluonbackx,\gluonbacky){4}
\end{picture}
\vskip 1cm
\caption{\label{I1234} Representative diagram for the type $I$ region 
of the four-gluon amplitude.}
\end{center}
\end{figure}
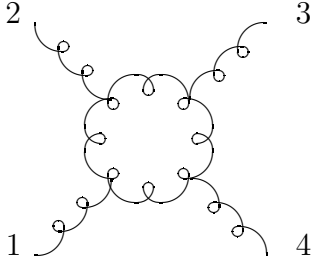

We start from region $I$, defined in this case by
\beq
\tau\to\infty\;\;,~~~~~~\hat\nu_{i+1,i}=O(1)~~~
{\rm for~all}~~1\leq i\leq 4~~~,
\label{reg14}
\eeq
where $\hat\nu_{54}\equiv 1-\hat\nu_4$.
Since we are looking for those contributions that in the field theory
limit diverge when $d\to 4$, again we need to isolate in the integrand 
of \eq{focus4} the term proportional to $\tau^{1-d/2}$.
Then, to find the divergence, we can replace the exponentials of the 
Green functions by a simple infrared cutoff, as we did in \eq{focus1}.
Substituting $\partial^2_\nu G(\nu)$ with the leading term
of its expansion, \eq{lartaud2G}, 
the contribution to $J(p_1,p_2,p_3,p_4)$
coming from region $I$ is
\bea 
J(p_1, p_2, p_3,p_4)\Big|_I & = & (d - 2) \, \ve_1\cdot
\ve_3\, \ve_2\cdot\ve_4 \int_0^\infty d\tau~
\tau^{3-d/2}\left(-\frac{2}{\tau}\right)^2    \nl
& \times & \int_0^1 d \hat\nu_4 \int_0^{\hat\nu_4} d \hat\nu_3 
\int_0^{\hat\nu_3} d \hat\nu_2~{\rm e}^{- 2 \a' m^2 \tau}  \nl
& = & {4\over 3} \, \ve_1\cdot\ve_3 \, \ve_2\cdot\ve_4
\, {1\over\e} \, + \, O(\e^0)~~~.
\label{4gfocus1}
\ena
We can check that this corresponds to the contribution
of the field theory diagrams with four
propagators in the loop, as depicted in \fig{I1234}.

\begin{figure}
\begin{center}
\begin{picture}(25000,8000)
\drawloop\gluon[\N 8](0,1100)
\drawline\gluon[\SW\REG](20,1500)[3]
\global\advance \gluonbackx by -1100
\put(\gluonbackx,\gluonbacky){1}
\drawline\gluon[\NW\REG](\gluonfrontx,\gluonfronty)[3]
\global\advance \gluonbackx by -1100
\put(\gluonbackx,\gluonbacky){2}
\drawline\gluon[\NE\REG](4000,3000)[2]
\global\advance \gluonbackx by 1100
\put(\gluonbackx,\gluonbacky){3}
\drawline\gluon[\SE\REG](4000,0)[2]
\global\advance \gluonbackx by 1100
\put(\gluonbackx,\gluonbacky){4}
\drawloop\gluon[\N 8](20000,1100)
\drawline\gluon[\SW\REG](21020,0)[2]
\global\advance \gluonbackx by -1100
\put(\gluonbackx,\gluonbacky){1}
\drawline\gluon[\NW\REG](21020,3000)[2]
\global\advance \gluonbackx by -1100
\put(\gluonbackx,\gluonbacky){2}
\drawline\gluon[\NE\REG](24800,1500)[3]
\global\advance \gluonbackx by 1100
\put(\gluonbackx,\gluonbacky){3}
\drawline\gluon[\SE\REG](\gluonfrontx,\gluonfronty)[3]
\global\advance \gluonbackx by 1100
\put(\gluonbackx,\gluonbacky){4}
\end{picture}
\vskip 1cm
\caption{\label{II'1234} 
Representative diagrams that do not contribute to the
divergence of the term proportional to $\varepsilon_1
\cdot\varepsilon_3\,\varepsilon_2\cdot\varepsilon_4$  
of the four-gluon amplitude. They instead contribute to the divergence 
of the term proportional to $\varepsilon_1
\cdot\varepsilon_2\,\varepsilon_3\cdot\varepsilon_4$.}
\end{center}
\end{figure}
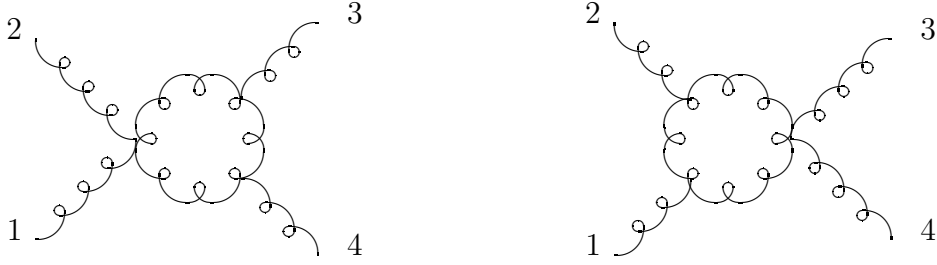

In the previous section we learnt that factors of $1/\tau$
are produced also by the type $II$ regions.
In the present case we have four such regions, namely 
\beq  
\tau\to\infty~~,~~~~~~\hat\nu_{i+1,i}=O(\tau^{-1})~~~,~~~1\leq i\leq4~~~.
\label{4greg2}
\eeq
To generate the desired factor of $1/\tau$ we must recover
in the integrand of \eq{focus4} an expression similar to
\eq{ktonu}, and thus we must employ 
the last two terms of the expansion (\ref{lartaud2G})
for both of the factors $\partial_\nu^2G(\nu)$ present in
\eq{focus4}. 
However, if only one pair of $\hat\nu$ variables satisfies
\eq{4greg2}, while the others are widely separated, only one factor
of $1/\tau$ is produced by the integration,
and the final result is not ultraviolet divergent in $d=4$. This means
that diagrams with three propagators in the loop,
like those depicted in \fig{II'1234}, do not contribute to 
the coefficient of the  $1/\e$ pole for the term 
$\ve_1\cdot\ve_3\,\ve_2\cdot\ve_4$.
On the contrary, to produce a divergence we must consider two of 
the regions in \eq{4greg2} at the same time. We have the 
following two possibilities
\beq  
\tau\to\infty~~,~~~\hat\nu_{2}=O(\tau^{-1})~~~{\rm and}~~~
\hat\nu_{43}=O(\tau^{-1})~~~,
\label{2fg}
\eeq
or
\beq  
\tau\to\infty~~,~~~\hat\nu_{32}=O(\tau^{-1})~~~{\rm and}~~~
1-\hat\nu_{4}=O(\tau^{-1})~~~.
\label{2fg2}
\eeq

\begin{figure}
\begin{center}
\begin{picture}(7000,8000)
\drawloop\gluon[\N 8](0,1100)
\drawline\gluon[\SW\REG](20,1500)[3]
\global\advance \gluonbackx by -1100
\put(\gluonbackx,\gluonbacky){1}
\drawline\gluon[\NW\REG](\gluonfrontx,\gluonfronty)[3]
\global\advance \gluonbackx by -1100
\put(\gluonbackx,\gluonbacky){2}
\drawline\gluon[\NE\REG](4800,1500)[3]
\global\advance \gluonbackx by 1100
\put(\gluonbackx,\gluonbacky){3}
\drawline\gluon[\SE\REG](\gluonfrontx,\gluonfronty)[3]
\global\advance \gluonbackx by 1100
\put(\gluonbackx,\gluonbacky){4}
\end{picture}
\vskip 1.5cm
\caption{\label{II(12)(34)} Representative diagram for the region 
$\hat\nu_{2}=O(\tau^{-1})$ and $\hat\nu_{43}=O(\tau^{-1})$
of the four-gluon amplitude.}
\end{center}
\end{figure}
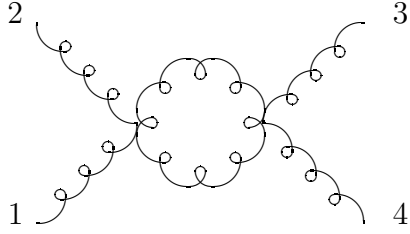

\begin{figure}
\begin{center}
\begin{picture}(7000,8000)
\drawloop\gluon[\N 8](0,1100)
\drawline\gluon[\NW\REG](2400,3900)[3]
\global\advance \gluonbackx by -1100
\put(\gluonbackx,\gluonbacky){2}
\drawline\gluon[\NE\REG](\gluonfrontx,\gluonfronty)[3]
\global\advance \gluonbackx by 1100
\put(\gluonbackx,\gluonbacky){3}
\drawline\gluon[\SW\REG](2400,-900)[3]
\global\advance \gluonbackx by -1100
\put(\gluonbackx,\gluonbacky){1}
\drawline\gluon[\SE\REG](\gluonfrontx,\gluonfronty)[3]
\global\advance \gluonbackx by 1100
\put(\gluonbackx,\gluonbacky){4}
\end{picture}
\vskip 1.5cm
\caption{\label{II(23)(41)} Representative diagram for the region 
$\hat\nu_{32}=O(\tau^{-1})$ and $1-\hat\nu_{4}=O(\tau^{-1})$
of the four-gluon amplitude.}
\end{center}
\end{figure}
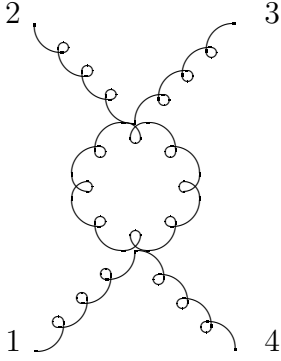

They correspond to the diagrams represented in \fig{II(12)(34)}
and \fig{II(23)(41)} respectively, which have two propagators and
two four-point vertices. The procedure is now almost identical to
the one followed in \secn{threepoint}. After using \eq{lartaud2G}, 
we find that the relevant terms of the form $k^{\hat\nu}$
combine to give  
\beq
16\, \left(k^{1+\hat\nu_2-\hat\nu_{43}}+
k^{1+\hat\nu_{43}-\hat\nu_2}\right)+\ldots
\label{fact1}
\eeq
in region (\ref{2fg}), and 
\beq
16\, \left(k^{\hat\nu_{32}+\hat\nu_{4}}+
k^{2-\hat\nu_4-\hat\nu_{32}}\right)+\ldots
\label{fact2}
\eeq
in region (\ref{2fg2}).
Computing the $\hat\nu$ integrals and adding the two
contributions, we find
\beq 
J(p_1, p_2, p_3,p_4) \Big|_{II} = - 16 \, 
\ve_1\cdot\ve_3 \,\ve_2\cdot\ve_4 \,
{1\over\e} \, + \, O(\e^0)~~~.
\label{4gtwodiv}
\eeq

\begin{figure}
\begin{center}
\begin{picture}(12000,10000)
\drawloop\gluon[\N 8](0,1100)
\drawline\gluon[\NE \REG](4000,3000)[1]
\drawline\gluon[\NW \REG](\gluonbackx,\gluonbacky)[3]
\global\advance \gluonbackx by -1100
\put(\gluonbackx,\gluonbacky){2}
\drawline\gluon[\NE\REG](\gluonfrontx,\gluonfronty)[3]
\global\advance \gluonbackx by 1100
\put(\gluonbackx,\gluonbacky){3}
\drawline\gluon[\SE\REG](\gluonfrontx,\gluonfronty)[3]
\global\advance \gluonbackx by 1100
\put(\gluonbackx,\gluonbacky){4}
\drawline\gluon[\SW \REG](1010,0)[2]
\global\advance \gluonbackx by -1100
\put(\gluonbackx,\gluonbacky){1}
\end{picture}
\vskip 1cm
\caption{\label{III1234} Representative diagram for the pinching  
$\nu_2\to\nu_3\to\nu_4$ in the four-gluon amplitude. 
The other three 
double pinchings, 
$\nu_3\to\nu_4\to\tau$, 
$\nu_2\to 0$ with $\nu_4\to\tau$, and $\nu_3\to\nu_2\to 0$, 
give rise to similar diagrams with the loop isolated 
respectively on the second, the third and the fourth external gluon.}
\end{center}
\end{figure}
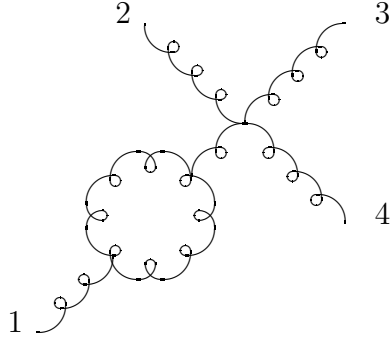

Finally, we turn our attention to one-particle reducible
contributions originating from the pinching regions. By analyzing the
structure of the integrand of \eq{focus4}, one realizes that
pinching singularities contribute to the ultraviolet divergence
of the term we are examining only when three consecutive punctures 
are pinched together, that is when
$\nu_2\to\nu_3\to\nu_4$, or $\nu_3\to\nu_4\to\tau$, or
$\nu_3\to\nu_2\to 0$, or $\nu_2\to 0$ with $\nu_4\to\tau$.
Each of these four regions gives the same contribution, that we 
expect to be simply proportional to the two-gluon amplitude, since
the loop is isolated on an external leg, as 
shown in \fig{III1234}.

Let us concentrate on the pinching $\nu_2\to\nu_3\to\nu_4$, which 
isolates the loop on the first leg. 
In this case \eq{focus4} becomes
\bea 
J(p_1, p_2, p_3,p_4) \Big|_{\nu_2\to\nu_3\to\nu_4} & = & - \, \ve_1\cdot
\ve_3\, \ve_2\cdot\ve_4 \int_0^\infty d\tau~
{\rm e}^{2\tau}~\tau^{-d/2}\nl
&\times&  \prod_{n=1}^\infty
\left(1-{\rm e}^{-2n\tau}\right)^{2-d}~
\int_0^\tau d  \nu_4\int_{\nu_4-\eta}^{\nu_4} d \nu_3   \nl 
& \times & \int_{\nu_3 - \eta}^{\nu_3} d \nu_2
\, {\rm e}^{2 \a'(p_2+p_3+p_4) \cdot p_1 
\, G(\nu_4)}\,\partial_4^2G(\nu_4)\nl
& \times & (\nu_{43})^{2\a' p_3\cdot p_4} (\nu_{32})^{2\a' p_2
\cdot p_3}(\nu_{42})^{-2+2\a' p_2\cdot p_4}  
\ena
where $\eta$ is an arbitrary small number.
By means of the change of variables 
\beq 
\nu_{3} \to x = \nu_{43}~~,~~~\nu_{2} \to y = {\nu_{32}\over\nu_{43}}~~~,
\label{declint}
\eeq
the integrals over $\nu_2$ and $\nu_3$ decouple and become 
respectively
\beq 
\int_0^\infty dy~(1 + y)^{- 2 + 2 \a' p_2 \cdot p_4}
y^{2 \a' p_2 \cdot p_3}=1+O(\a')~~~,
\label{4gint2}
\eeq
and
\beq 
\int_0^\eta dx~x^{-1 + 2 \a'(p_2 \cdot p_3 + p_3
\cdot p_4 + p_2 \cdot p_4)}~~~. 
\label{4gint1}
\eeq 
Notice that this last integral is of the form of \eq{pinchser}
with $n=1$. Then, for $\a'\to 0$ we get
\beqa
J(p_1, p_2, p_3,p_4) \Big|_{\nu_2\to\nu_3\to\nu_4} 
& = & \ve_1\cdot\ve_3 \, \ve_2\cdot
\ve_4 \, {(p_2 + p_3 + p_4) \cdot p_1 \over p_2 \cdot p_3 +
p_3 \cdot p_4 + p_2 \cdot p_4} \nl
& \times & R \left((p_2 + p_3 + p_4) \cdot p_1 \right) ~~~.
\label{4gamb}
\eeqa
As expected, this last equation contains a 
ratio of invariants that becomes
indeterminate when the external states are on shell.
We solve the ambiguity with the same prescription used in the
previous section, {\it i.e.} we put off shell only the momentum 
of the gluon attached to the loop and keep the others on shell, 
according to
\beq 
p_1^2 = m^2~~~,~~~~~~p_2^2=p_3^2=p_4^2=0~~~.
\label{4goffs}
\eeq
Using momentum conservation and \eq{4goffs}, we easily see that
\beq 
{(p_2+p_3+p_4) \cdot p_1 \over p_2 \cdot p_3 +
p_3 \cdot p_4 + p_2 \cdot p_4} = - 2~~~. 
\label{4gsolv}
\eeq
Hence
\bea
J(p_1, p_2, p_3,p_4) \Big|_{\nu_2\to\nu_3\to\nu_4}  &=& -2\,R(-m^2)\,
\ve_1\cdot\ve_3\, \ve_2\cdot\ve_4 \nl
&=&\frac{44}{3}\,\ve_1\cdot\ve_3\, \ve_2\cdot\ve_4 \,\frac{1}{\epsilon} 
+O(\epsilon^0)~~~,
\label{4gpinch}
\ena
where again we have used \eq{Rfin} for $d=4-2\e$.
The other three pinchings can be analyzed in a 
similar way and lead to the same result of \eq{4gpinch}. Therefore,
the total contribution to $J(p_1,p_2,p_3,p_4)$ from region $III$ is
\beq
J(p_1,p_2,p_3,p_4)\Big|_{III} = 4 \, 
\frac{44}{3}\,\ve_1\cdot\ve_3\, \ve_2\cdot\ve_4 \,\frac{1}{\epsilon} 
+O(\epsilon^0)~~~.
\label{4gtotpinch}
\eeq

Collecting all the results derived so far, and
reinstating all normalization factors, we find
that
\beq 
A^{(1)}{[13,24]}\Big|_{\rm div} = 
3N\,\frac{g^2}{(4\pi)^2}\,\frac{11}{3}\,\frac{1}{\e}
\, A^{(0)}{[13,24]}~~~,
\label{4gtotdiv}
\eeq
where the symbols $A^{(1)}{[13,24]}$ and $A^{(0)}{[13,24]}$
denote the terms proportional to $\ve_1\cdot\ve_3\, \ve_2\cdot\ve_4$
in the four-gluon amplitude at one loop and tree level respectively.
Comparing with what we expect from field 
theory for a truncated, connected four-point
Green function, we obtain
\beq 
Z_4= 1 + N\,\frac{g^2}{(4\pi)^2}\,\frac{11}{3}\,\frac{1}{\e}~~~.
\label{z4}
\eeq
Thus, we are led to the background field Ward identities
\beq
Z_A=Z_3=Z_4~~~,
\label{z4=za}
\eeq
as desired.

This result is further confirmed by the fact that the contributions
from the type $I$ and type $II$ regions, associated with one-particle 
irreducible diagrams, properly add. Indeed,
\bea
A^{(1)}{[13,24]}\Big|_{\rm 1PI} &\equiv&
A^{(1)}{[13,24]}\Big|_{I} + A^{(1)}{[13,24]}\Big|_{II} 
\nonumber \\
&=&
- N\,\frac{g^2}{(4\pi)^2}\,\frac{11}{3}\,\frac{1}{\e}
\, A^{(0)}[13,24]
+ O(\e^0)~~~,
\label{1pi4}
\ena
which leads directly to \eq{z4}.

We conclude with a brief analysis of the term proportional 
to $\ve_1\cdot\ve_2\,\ve_3\cdot\ve_4$ in the four-gluon amplitude, and 
show that our results, \eq{z4} and \eq{z4=za}, still hold in this case.
Let us then consider the integral
\bea
{\tilde J}(p_1, p_2, p_3,p_4) & = & 
\int_0^\infty d\tau~{\rm e}^{2\tau} \,
\tau^{-d/2} \prod_{n=1}^\infty
\left(1-{\rm e}^{-2n\tau}\right)^{2-d}\int_0^\tau d  \nu_4
\int_0^{\nu_4} d\nu_3 \int_0^{\nu_3} d\nu_2  \nl
& \times & {\rm e}^{2\a'p_1\cdot p_2 \, G(\nu_{2})}~
{\rm e}^{2\a'p_1\cdot p_3 \, G(\nu_{3})}~
{\rm e}^{2\a'p_1\cdot p_4 \, G(\nu_{4})}~\label{focus44}\\
& \times &
{\rm e}^{2\a'p_2\cdot p_3 \, G(\nu_{32})}~ 
{\rm e}^{2\a'p_2\cdot p_4 \, G(\nu_{42})}~
{\rm e}^{2\a'p_3\cdot p_4 \, G(\nu_{43})} \nl
& \times &\ve_1\cdot\ve_2 \,\ve_3\cdot\ve_4 \,
\partial_2^2 G(\nu_2)
\partial_4^2 G(\nu_{43})~~~,
\nonumber
\ena
and let us compute its field theory limit. The contribution from the 
type $I$ region, \eq{reg14}, is as before, namely
\beq 
{\tilde J}(p_1, p_2, p_3, p_4)\Big|_I  = 
 {4\over 3} \, \ve_1\cdot\ve_2 \, \ve_3\cdot\ve_4
\, {1\over\e} \, + \, O(\e^0)~~~.
\label{4gfocus14}
\eeq

The contribution from the type $II$ region is, however, different
from \eq{4gtwodiv}. This can be traced to the fact that in this case there
are contributions from regions of type $II$ in which the two $\hat\nu$ 
variables that are taken to be close to each other are precisely the ones
that appear in the argument of one of the Green functions whose double 
derivative is taken in \eq{focus44}. As a consequence, there will now 
be non-vanishing contributions also when a single 
pair of $\hat\nu$ variables approach each other. Let us, for instance, 
consider the region
\beq
\tau \rightarrow \infty ~~~,~~~~~~
\hat\nu_{2} = O(\tau^{-1})~~~.
\label{reg24}
\eeq
with all other variables widely separated. It corresponds to the diagram 
depicted on the left in \fig{II'1234}. Using \eq{lartaud2G}, we see that 
in this region the relevant terms of 
$\partial_2^2G(\nu_2)\,\partial_4^2G(\nu_{43})$
combine to give
\beq
\frac{8}{\tau}\,\left(
\sum_{n=1}^\infty n\,{\rm e}^{-2n\tau{\hat\nu}_2}
+{\rm e}^{- 2 \tau (1 - {\hat\nu}_2)}
+{\rm e}^{- 2 \tau (1 + {\hat\nu}_2)}\right)+\ldots~~~.
\label{d2G2}
\eeq
Now however, in contrast to all previous cases, the entire series
in $n$, and not just its first term, must be taken into account: in fact,
in region (\ref{reg24})
all terms of the form ${\rm e}^{-2n\tau{\hat\nu_2}}$ 
are on equal footing for any $n$ since $\tau{\hat\nu}_2$
is finite.
Computing the integrals, and extracting the right power of 
$\tau$ that is necessary to produce the divergence in $d=4-2\epsilon$, 
we find that
\bea
{\tilde J}(p_1, p_2, p_3, p_4)\Big|_{\hat\nu_{2} = O(\tau^{-1})}  &=&
4\left(\sum_{n=1}^\infty 1\right) \ve_1\cdot\ve_2 \, \ve_3\cdot\ve_4
\, {1\over\e} \, + \, O(\e^0) \nl
&=& -2\,\ve_1\cdot\ve_2 \, \ve_3\cdot\ve_4
\, {1\over\e} \, + \, O(\e^0) 
\label{4gfocus24}
\ena
where we have regularized the divergent sum $\sum\limits_{n=1}^\infty 1$
by means of the formula \footnote{This $\zeta$-function 
regularization will be used also in \secn{effact}, following 
\cite{mets}.}
\beq
\sum_{n=1}^{\infty} 1 = \lim_{s \rightarrow 0} 
\sum_{n=1}^{\infty} n^{-s}=
\zeta (0) = - \frac{1}{2}~~~.
\label{zreg}
\eeq
It should be apparent that the reason for this divergence is the singularity
of the Green function as $\nu_2 \to 0$, which did not cause problems when 
only the first term in the series in \eq{lartaunuG} was used, but cannot be 
avoided when the whole series is included. One may thus wonder whether we are
not double counting some terms, which might as well be included in the 
pinching region. This does not happen, and the reason is that the residual
contribution from the pinching region, which is included in \eq{4gfocus24},
can be traced to the exchange of a tachyon in the pinched channel,
that is explicitly discarded when we study region $III$. In fact, if
one uses \eq{pinchpol} for $n = 0$ to isolate the tachyon contribution in the 
relevant channel, the result is divergent in the field theory limit.
This divergence can be regularized by taking the residue of the tachyon pole,
which corresponds to a contact interaction with a four-gluon vertex and
no propagating tachyon. This calculation leads precisely to a contribution
matching \eq{4gfocus24}. By including this contact interaction in region $II$
we are correctly identifying it as a one-particle irreducible contribution.

A similar reasoning leads to identify an identical contribution 
from the region
\beq
\tau \rightarrow \infty ~~~,~~~~~~
\hat\nu_{43} = O(\tau^{-1})~~~,
\label{reg243}
\eeq
which corresponds to the diagram depicted on the right of \fig{II'1234}.

Finally, we must consider the contribution from regions (\ref{2fg}) 
and (\ref{2fg2}), associated with the diagrams in \fig{II(12)(34)}
and \fig{II(23)(41)} respectively.
The relevant terms of $\partial_2^2G(\nu_2)\,\partial_4^2G(\nu_{43})$
combine to give
\bea
&&\!\!\!\!\!\!\!\!\!16\,\left(
\sum_{n=1}^\infty n\,{\rm e}^{-2n\tau{\hat\nu}_2}
+{\rm e}^{- 2 \tau (1 - {\hat\nu}_2)}
+{\rm e}^{- 2 \tau (1 + {\hat\nu}_2)}\right)\nl
&\times&\left(
\sum_{m=1}^\infty m\,{\rm e}^{-2m\tau{\hat\nu}_{43}}
+{\rm e}^{- 2 \tau (1 - {\hat\nu}_{43})}
+{\rm e}^{- 2 \tau (1 + {\hat\nu}_{43})}\right)+\ldots~~~,
\label{d2G24}
\ena
in region (\ref{2fg}), while in region ({\ref{2fg2}) they give
\beq
16\,\left(
{\rm e}^{-2\tau\left({\hat\nu}_4-{\hat\nu}_{32}\right)}
+{\rm e}^{- 2 \tau (2 - {\hat\nu}_4+{\hat\nu}_{32})}
\right)+\ldots~~~.
\label{d2G244}
\eeq

Computing the integrals we find that, after using twice the $\zeta$-
function regularization, \eq{d2G24} yields
\beq
2\,\ve_1\cdot\ve_2 \, \ve_3\cdot\ve_4
\, {1\over\e} \, + \, O(\e^0) ~~~,
\label{cara1}
\eeq
while \eq{d2G244} yields
\beq
8\,\ve_1\cdot\ve_2 \, \ve_3\cdot\ve_4
\, {1\over\e} \, + \, O(\e^0) ~~~.
\label{cara2}
\eeq
Adding all contributions of the type $II$ regions, we see that 
\beq 
{\tilde J}(p_1, p_2, p_3, p_4)\Big|_{II}  = 
 6 \, \ve_1\cdot\ve_2 \, \ve_3\cdot\ve_4
\, {1\over\e} \, + \, O(\e^0)~~~.
\label{4gfocus241}
\eeq

As we have discussed above, the regions of type $I$ and $II$ are 
associated with diagrams that are 1PI. Then, adding the contributions
of these regions, \eq{4gfocus14} and \eq{4gfocus241},
and reinstating all normalization factors, we get, in obvious
notations,
\bea
A^{(1)}{[12,34]}\Big|_{\rm 1PI} &\equiv&
A^{(1)}{[12,34]}\Big|_{I} + A^{(1)}{[12,34]}\Big|_{II} 
\nl
&=&
- N\,\frac{g^2}{(4\pi)^2}\,\frac{11}{3}\,\frac{1}{\e}
\, A^{(0)}[12,34]\Big|_{\rm 1PI}
+ O(\e^0)~~~,
\label{1pi41}
\ena
where $A^{(0)}[12,34]\Big|_{\rm 1PI}$ stands for
the color-ordered tree-level 
1PI term proportional to $\ve_1\cdot\ve_2\,\ve_3\cdot\ve_4$ 
in the Feynman gauge, that is 
\beq
A^{(0)}[12,34]\Big|_{\rm 1PI} = 
-2~g_d^2~{\rm Tr}
(\lambda^{a_1}\lambda^{a_2}\lambda^{a_3}\lambda^{a_4})\,
\ve_1\cdot\ve_2\,\ve_3\cdot\ve_4 ~~~.
\label{1pi40}
\eeq
Clearly, \eq{1pi41} leads to the correct value of $Z_4$, \eq{z4}. 

Finally, the pinching contributions to ${\tilde J}(p_1,p_2,p_3,p_4)$ due
to gluon exchanges can be easily computed along the lines discussed 
above. If we also add them, we get
\beq 
A^{(1)}{[12,34]}\Big|_{\rm div} = 
3N\,\frac{g^2}{(4\pi)^2}\,\frac{11}{3}\,\frac{1}{\e}
\, A^{(0)}{[12,34]}~~~,
\label{4gtotdiv1}
\eeq
as expected. 

\sect{A direct computation of proper vertices}
\label{effact}
\vskip 0.5cm

{}From the detailed analysis of the previous sections we learnt that it is 
possible, and indeed desirable, if one wishes to calculate renormalization 
constants, to isolate from the amplitude one-particle irreducible diagrams 
that contribute to the effective action. 
In this section we will show how this can be done
directly at the level of the string amplitude, following the ideas of
Metsaev and Tseytlin~\cite{mets}. The starting point of this approach 
is the introduction of an alternative expression for the bosonic Green
function, \eq{newmtgreen}, which is suitable for a regularization of
pinching singularities. As we shall see, the $\zeta$-function regularization
used in this case is precisely the one introduced in \eq{zreg}, and thus
it handles correctly also the residual contributions from the propagation 
of the tachyon in the pinched channel, assigning them to one-particle 
irreducible diagrams. The most interesting feature of this approach
is that, at least at one loop, it allows to integrate exactly over the 
punctures, before the field theory limit is taken. This is possible because,
having regularized the pinching singularity, there is no integration
region that generates negative powers of $\a'$ in the field theory limit.
One can then focus on terms proportional to $(\a')^{2 - d/2}$, and for those
terms one is allowed to replace the exponentials of the Green function
by a simple infrared cutoff, of the form ${\rm exp}(- 2 \a' m^2 \tau)$,
as we learnt in the previous sections. The simplified integral over the
punctures can then be performed exactly, using the Green function 
given by \eq{newmtgreen}. Using these techniques we will be able to express
the renormalizations of the various divergent amplitudes in terms of a single
string integral, $Z(d)$, which in the field theory limit reproduces the
wave function renormalization $Z_A$.

Our starting point is \eq{onemaster}, with the choice of projective 
transformations given in \eq{gaugech}. We rewrite it as
\beqa
A^{(1)}_P (p_1,\ldots,p_M) & = & N \, {\rm Tr}(\lambda^{a_1}
\cdots \lambda^{a_M}) \,
\frac{g_d^M}{(4\pi)^{d/2}} \, (2 \a')^{2 - d/2} \nl 
& \times & (- 1)^M \, \int_{0}^{\infty} {\cal D}\tau \,
I^{(1)}_M(\tau)~~~,
\label{newonemast}
\eeqa
where
\beqa
I^{(1)}_M(\tau) & = & (2 \a')^{M/2 -2}
\int_{0}^{\tau} d \nu_{M} \int_{0}^{\nu_{M}} d \nu_{M-1} \dots 
\int_{0}^{\nu_3} d \nu_2 \left\{\prod_{i<j} 
\left[ \exp\left(G(\nu_{ji})\right) \right]^{2\a' p_i\cdot p_j} 
\right.  \nl
& \times & \left. \exp \left[ \sum_{i\not=j}
\left(\sqrt{2\a'} p_j\cdot\ve_i 
\, \partial_i  G(\nu_{ji}) + {1\over 2} 
\ve_i\cdot\ve_j \, \partial_i\partial_j
G(\nu_{ji}) \right) \right] \right\}_{\rm m.l.}~~~.
\label{Mint}
\eeqa 
 
For $M=2$, after a partial integration with vanishing surface term, we get
\beq
I^{(1)}_2(\tau) = \ve_1 \cdot \ve_2 \,p_1 \cdot p_2   
\int_{0}^{\tau} d \nu
\Big( \partial_\nu G(\nu) \Big)^2 {\rm e}^{2\a'p_1\cdot p_2 \, G(\nu)}~~~.
\label{2int}
\eeq
As announced, since the overall power of $\a'$ is already appropriate, 
we can now neglect the exponential, and replace it with an infrared
cutoff to isolate ultraviolet divergences.
Using the expression in \eq{newmtgreen} for the Green 
function, we can easily perform exactly the integral over the puncture, 
by means of the identity
\beq
\int_{0}^{1} dx\, \sin (2 \pi n x) \, \sin (2 \pi mx) = 
\frac{1}{2} \, \delta_{nm}~~~.
\label{sinsin}
\eeq
 
We get
\beq
I^{(1)}_2(\tau) = \frac{2 \pi^2}{\tau} \, \ve_1 \cdot \ve_2 \,p_1 \cdot p_2
\, \sum_{n=1}^{\infty} \left(
\frac{1 + q^{2n}}{1 - q^{2n}} \right)^2~~~,
\label{2res}
\eeq
which allows one to write
\beqa
A^{(1)}(p_1, p_2) \Big|_{\rm div} & = & 
\frac{N}{2}\, {\rm Tr} \left( \lambda^{a_1}
\lambda^{a_2} \right) 
\frac{g_d^2}{(4 \pi)^{d/2}} (2 \a')^{2 - d/2}  
\, \ve_1 \cdot \ve_2 \,p_1 \cdot p_2\,Z(d)  \nl
& = & \frac{N}{4} \, \frac{g_d^2}{(4 \pi)^{d/2}} \, ( 2 \alpha ')^{2 -d/2} 
\, Z(d) \, A^{(0)}(p_1, p_2)~~~.
\label{2onetree}
\eeqa
Here
\beq
Z(d) \equiv (2 \pi)^2 \int_0^\infty \frac{{\cal D} \tau}{\tau} \,
\sum_{m=1}^{\infty}
\left(\frac{1 + q^{2m}}{1- q^{2m}}\right)^2
\label{Zint}
\eeq
is the string integral that generates the renormalization constants as 
$\a' \to 0$, and, with our normalizations
\beq
A^{(0)}(p_1, p_2) = 2 \, {\rm Tr} (\lambda^{a_1} \lambda^{a_2}) \,
\ve_1 \cdot \ve_2 \,p_1 \cdot p_2~~~.
\label{2tree}
\eeq
Since the integral $Z(d)$ will reappear in the analysis of the three
and four-point amplitudes, we postpone its evaluation and go on to
perform the integral over the punctures for the other amplitudes.

With three gluons we get
\beqa
I^{(1)}_3(\tau) & = & \int_{0}^{\tau} d \nu_3 \int_{0}^{\nu_3} 
d \nu_2 \Bigg\{- \ve_1 \cdot \ve_2 \, \partial_2^2 
G(\nu_{2})  \nl
& \times &  \Big[ p_1 \cdot \ve_3 \, \partial_3 G (\nu_{3}) +
p_2 \cdot \ve_3 \, \partial_3 G(\nu_{32}) \Big] 
+ \dots \Bigg\}~~~, 
\label{3int}
\eeqa
where terms needed for cyclic symmetry and terms of order $\a'$ are not 
written explicitly, and we discarded the exponentials of the Green
functions, that are again irrelevant for ultraviolet divergences.
Notice that for the three-point function the overall power of $\a'$
is correct provided we do not perform the partial integration of double 
derivatives, as was done for the two-point amplitude. As remarked at the end
of \secn{threepoint}, partial integration of the three-gluon integrand
reshuffles the contributions of different diagrams in the field theory limit,
so that it is no longer possible to identify unambiguously the one-particle
irreducible ones.
 
The integrals over $\nu_2$ and $\nu_3$ can now be done, using as before 
\eq{newmtgreen} for the Green function. It is easier to start 
by performing the integral over $\nu_3$, from $\nu_2$ to $\infty$. Then 
the integral over $\nu_2$ can be performed, as in the case of the two-gluon 
amplitude, using an identity similar to \eq{sinsin}, but with the cosines
replacing the sines. The result is 
\beqa
I^{(1)}_3(\tau) & = & \frac{(2 \pi)^2}{\tau}
\Big[\ve_1 \cdot \ve_2\, p_2 \cdot \ve_3 +
\ve_2 \cdot \ve_3\, p_3 \cdot \ve_1 +
\ve_1 \cdot \ve_3\, p_1 \cdot \ve_2 \Big]  \nl
& \times & \sum_{n=1}^{\infty} 
\left( \frac{1+ q^{2n}}{1 - q^{2n}} \right)^2 \, + \, O(\a')~~~,
\label{3res}
\eeqa
so that the three-gluon amplitude is given by
\beq
A^{(1)}(p_1, p_2, p_3) \Big|_{\rm div} = \frac{N}{4} 
\frac{g_d^2}{(4 \pi)^{d/2}} (2 \a')^{2 - d/2}\, Z(d)\, A^{(0)}(p_1, p_2, p_3)
\, + \, O(\a')~~~,
\label{3onetree}
\eeq
and is once again expressed in terms of the integral $Z(d)$.
When the integral $Z(d)$ is evaluated (as done explicitly in Appendix B),
and the pinching singularities regularized, \eq{3onetree} will reduce to
\eq{1pi3}, as expected.

For completeness, we now show how this procedure goes through for the 
four-point amplitude. In this case we can concentrate on those terms 
in the amplitude that have no powers of the external momenta (and thus have 
a coefficient of the type $\ve_i \cdot \ve_j \, \ve_h \cdot \ve_k$). 
These have the correct power of $\a'$ before partial integration, so 
we can again discard the exponentials. 
Then we need to compute
\beqa
I^{(1)}_4(\tau) & = & \int_0^\tau d \nu_4 \int_0^{\nu_4} d \nu_3 
\int_0^{\nu_3} d \nu_2 \,\Bigg[ \ve_1 \cdot \ve_2 \, 
\ve_3 \cdot \ve_4 \, \partial_2^2 G(\nu_{2}) \, 
\partial_4^2 G(\nu_{43})  \nl
& + & \ve_1 \cdot \ve_3  \, 
\ve_2 \cdot \ve_4 \, \partial_3^2 G(\nu_{3}) \,
\partial_4^2 G(\nu_{42})  \label{4int}  \\
& + & \ve_1 \cdot \ve_4  \, 
\ve_3 \cdot \ve_2 \, \partial_4^2 G(\nu_{4}) \,
\partial_3^2 G( \nu_{32}) \, + \, \ldots \Bigg]~~~.
\nonumber
\eeqa
 
Using again \eq{newmtgreen}, we can perform the integrals over the punctures,
and we get
\beq
I^{(1)}_4(\tau) = \frac{(2 \pi)^2}{\tau} \sum_{n=1}^{\infty} \left(\frac{1 + 
q^{2n}}{1 - q^{2n}} \right)^2 \left[ - \frac{1}{2}
\ve_1 \cdot \ve_2  \, 
\ve_3 \cdot \ve_4  +
\ve_1 \cdot \ve_3  \,
\ve_2 \cdot \ve_4 - \frac{1}{2}
\ve_2 \cdot \ve_3  \, 
\ve_1 \cdot \ve_4 \right]~~~.
\label{4res}
\eeq
The amplitude becomes then
\beq
A^{(1)}_P(p_1, p_2, p_3, p_4) \Big|_{\rm div} = \frac{N}{4} 
\frac{g_d^2}{(4 \pi)^{d/2}} (2 \alpha ')^{2 - d/2}\, Z(d)\, 
A^{(0)}(p_1, p_2, p_3, p_4) \, + \, O(\a')~~~,
\label{4onetree}
\eeq
where the 1PI part of the tree-level amplitude in Feynman gauge
has the structure given in the square bracket of \eq{4res}, as can
easily be checked.
Once again, the divergent part of the one-loop
amplitude is expressed in terms of the basic integral $Z(d)$. 
The previous results for the tree-level and one-loop proper vertices 
can be obtained from the effective Lagrangian
\beq
L = - \frac{1}{4} \, F_{\mu \nu}^{a}  F^{\mu \nu}_{a} \, 
\Big[1 + K(d) \Big]~~~,
\label{efflag}
\eeq
where
\beq
K(d) = \frac{N}{4} \,
\frac{g_d^2}{(4 \pi)^{d/2}} \, (2 \a')^{2 -d/2} \,  Z(d)~~~.
\label{loopefflag}
\eeq
If we now perform the limit $\a' \to 0$, keeping the ultraviolet 
cutoff $\e \equiv 2 - d/2$ small but positive, and eliminating by hand 
the tachyon contribution, as described in detail in Appendix B, we arrive at
\beq
K(4 - 2 \e) \rightarrow
- \frac{11}{3} \, N \, \frac{g^2}{(4\pi)^2} \, \frac{1}{\e} \,
+ \, O(\e^0)~~~.
\label{divefflag}
\eeq
  
We see that one-loop diagrams generate divergences that can be eliminated 
by the addition of a counterterm of the form
\beq
L_{c.t.} = - \frac{1}{4} \, F_{\mu \nu}^{a} F^{\mu \nu}_{a} \, K_A~~~,
\label{efflagct}
\eeq
where $K_A = - K$.
This implies that the wave function renormalization constant is
\beq
Z_A \equiv 1 + K_A = 1   + \frac{11}{3} \, N \, \frac{g^2}{(4 \pi)^2} \, 
\frac{1}{\e}~~~,
\label{zaagain}
\eeq
in agreement with \eq{za}. It also implies
\beq
Z_A = Z_3 = Z_4~~~,
\label{wardagain}
\eeq
in agreement with the results of the previous sections, and as expected in
the background field method.

\sect{String in a non-abelian background field}
\label{background}
\vskip 0.5cm

The results of the previous sections show that the field theory limit of 
string theory leads unambiguously to the background field method. 
Furthermore, we have learnt that contributions to 
the effective action can be isolated from 
the amplitudes by regularizing the pinching singularities in the Green 
function. In this section we will show how, under appropriate assumptions,
these results can be derived directly, for any number of string loops, 
from the partition function of a string interacting with an external 
non-abelian background field. We will give a general expression for 
the planar contribution to this partition function, and we will formally 
generalize the analysis of \secn{effact} to all loops, verifying explicitly 
that gauge invariance is respected, and constructing a general expression
for the multiloop analog of the one-loop integral $Z(d)$.
We will then deal with the special problems arising at 
tree level and one loop, along the lines of Ref.~\cite{tselett}, and show how 
the results of the previous sections are correctly reproduced with this
construction. This also substantiates the somewhat formal manipulations 
that one has to perform on the multiloop expressions.

Let us consider the planar contribution to the partition function of an open 
bosonic string interacting with an external non-abelian $SU(N)$ background.
It is given by 
\beqa
Z_{P}(A_{\mu}) & = & \sum_{h=0}^{\infty} N^h g_s^{2 h -2} 
\int DX~ Dg~ 
{\rm e}^{- S_0(X,g;h)} \times \nl
& \times & {\rm Tr} \left[ P_z \exp \left(  {\rm i}  g_d  
\int_h dz ~\partial_z X^{\mu} (z) A_{\mu} ( X (z))    
\right) \right]~~~. 
\label{part}
\eeqa
The path-ordering symbol $P_z$ reminds us that in the open 
string the $z$ variables are ordered, as in \eq{cyclord}, 
along the world-sheet boundary; it is defined by
\beqa
& & {\rm Tr} \left[ P_z \exp \left(  {\rm i}  g_d  
\int_h dz ~\partial_z X^{\mu} (z) A_{\mu} ( X (z))    
\right) \right] 
=  \sum_{n=0}^{\infty} \left( {\rm i} g_d  \right)^n 
\label{pathordered} \\
& \times &  \int_{\Gamma_{h,n}} \prod_{i=1}^{n} d z_i ~\partial_{z_1} 
X^{\mu_1} (z_1 )
\dots \partial_{z_n} X^{\mu_n} (z_n )~{\rm Tr} 
\left[ A_{\mu_1} (X(z_1 )) 
\dots A_{\mu_n} (X(z_n ))\right]~~~.
\nonumber 
\eeqa 
The precise region of integration for the punctures $z_i$ will in general 
depend on the moduli of the open Riemann surface, and we denoted it by 
$\Gamma_{h,n}$, for a surface of genus $h$ with $n$ punctures. The gauge 
coupling constant $g_d$ and the dimensionless string coupling $g_s$ are 
related by \eq{gstring}. Finally the bosonic string action on a genus $h$ 
manifold with world sheet metric $g_{\alpha \beta}$ is  
\beq
S_0(X,g;h) = \frac{1}{4 \pi \a'} 
\int_h d^2 z ~\sqrt{g} g^{\alpha \beta} \partial_{\alpha} X (z) 
\cdot  \partial_{\beta} X(z)~~~.
\label{classact}
\eeq

It is convenient to separate the zero mode $x^{\mu}$ in the string 
coordinate $X^{\mu}$, defining
\beq
X^\mu(z) = x^\mu +  (2 \a' )^{1/2} \xi^\mu(z)~~,
\label{xmu}
\eeq
so that $\xi^\mu$ is dimensionless, while the zero mode $x^\mu$ as well as 
the string coordinate $X^{\mu}$ have dimensions of length. In terms of
$x^{\mu}$ and $\xi^{\mu}$ the measure of the functional integral in 
\eq{part} becomes
\beq
D X = \frac{d^d x}{(2 \a' )^{d/2}} D \xi~~~.
\label{measu}
\eeq 

Inserting in \eq{part} the representation of the path-ordered 
exponential given by \eq{pathordered},  and concentrating on
terms up to $O(2 \a')^2$, we can write the partition function as
\beqa
Z_{P}(A_{\mu}) & = & 
\sum_{h=0}^{\infty} \left(\frac{N}{(2 \pi)^d}\right)^h 
g_{s}^{2h-2} \int 
\frac{d^d x}{(2 \a ')^{d/2}} \int d {\cal{M}}_h \Bigg\{ {\rm Tr} (1) \nl
& - & g_{d}^{2} \left[ C^{(h)}_2(A) + {\rm i} g_d C^{(h)}_3(A) + 
({\rm i} g_d )^2  C^{(h)}_4(A) \right] \Big\}~~~,
\label{expansion}
\eeqa
where $C^{(h)}_i(A)$ corresponds to the 
contributions of terms with $i$ external gauge fields, and is obtained 
performing the functional integral over the variable $\xi$. 

The measure of
integration over the moduli, $d {\cal{M}}_h$, is equal to the one given in 
\eq{hmeasure}, but does not include the differentials of the punctures, 
nor at this stage the factor $dV_{abc}$, responsible for the fixing of 
projective invariance. For $h \geq 2$ the factor $dV_{abc}$ can be included
without problems, provided we do not fix the position of any of the punctures,
choosing instead three of the fixed points of the surface in the Schottky 
representation. For $h = 0$ this cannot be done, and for $h = 1$ it can be
done only partially, as the surface in these cases has less than three 
fixed points. One is then led to fix some of the punctures, as was done 
in the previous sections. In the present context, however, fixing some of the
punctures would interfere with the definition of path ordering given by
\eq{pathordered}, and would significantly complicate the following derivation.
We then choose not to fix any of the punctures, which ensures that our 
formulas are valid for any number of loops. The price we pay is that at 
tree level the expressions we write are formally infinite, 
because we failed to divide by the volume of the projective group. 
This however can be repaired by treating the projective infinities 
with a renormalization prescription: first we regularize them 
by compactifying the range of integration, then we divide by the compactified 
projective volume, obtaining a quantity that remains
finite when the compactification radius is taken to infinity. As with
any renormalization prescription, this leaves us 
with an undetermined finite overall constant, that can be fixed by comparing
with the results of the previous sections, or with a field theory calculation.
For this reason we are unable to deduce \eq{gstring} in this functional 
integral approach (see also \cite{tselett}).
At one loop, we can partially deal with the projective invariance by 
fixing the positions of the two fixed points of the Schottky generator.
We are then left with an overall translational invariance, that we do
not eliminate by fixing one of the punctures. Rather, we integrate over 
all punctures, dividing at the end by the volume of translation, which 
is finite at one loop. This way we get the correct one-loop result, 
including the overall constant.

Since \eq{part} is the string expectation value of a Wilson 
loop winding around one boundary of the string, we expect it 
to be gauge invariant. It is 
instructive to see how gauge invariance emerges out of the calculation, for
any number of string loops. To this end, we will compute separately the three 
contributions $C^{(h)}_{2}(A), C^{(h)}_{3}(A)$ and $C^{(h)}_4(A)$, and 
see how they correctly reconstruct the gauge invariant effective action. 
We will assume that pinching singularities
have been regularized, and thus we will discard all terms arising from 
partial integrations that involve the Green function evaluated at 
the pinching. We will also use the periodicity of the Green function 
to discard all boundary terms corresponding to integrations over the 
complete boundary loop.

Let us start by considering $C^{(h)}_2(A)$. It is given by
\beq
C^{(h)}_2(A) = \left< \frac{1}{2} \int dz_1 \int dz_2 ~\theta(z_1 -z_2 ) 
\partial_{z_1} X^{\mu} (z_1) \partial_{z_2} X^{\nu} (z_2)
A^{a}_{\mu} ( X(z_1) ) A^{a}_{\nu} ( X(z_2) )  \right>~~~,
\label{c2}
\eeq
where the expectation value is given by a functional integration over the 
field $\xi$.

Since the integrand (apart from the $\theta$-function) in \eq{c2}
is symmetric under the exchange of the indices $1$ and $2$ we can also write 
\beq
C^{(h)}_2(A) = \frac{1}{4}\left<  \int dz_1 \int dz_2 ~
\partial_{z_1} X^{\mu} (z_1) \partial_{z_2} X^{\nu} (z_2)
A^{a}_{\mu} ( X(z_1) ) A^{a}_{\nu} ( X(z_2) )  \right>~~~,
\label{c2'}
\eeq 
where now $z_1$ and $z_2$ are integrated independently in the same domain.

If we now expand the gauge field around the string center of mass
using \eq{xmu}, we include terms up to order $(2  \a' )^2 $, and we
discard total derivatives, we get
\beqa
C^{(h)}_2(A) & = & \frac{(2 \a')^{3/2}}{4} \Bigg< \int  d z_1 
\int dz_2 ~ 
\partial_{z_1} \xi^\mu (z_1) \partial_{z_2} \xi^\nu(z_2) \nl
& \times & \Big[ \xi^{\rho}(z_1)  \partial_{\rho} A_{\mu}^{a}(x)
A^{a}_{\nu} (x) + \xi^{\rho}(z_2)  \partial_{\rho} A_{\nu}^{a} (x) 
A^{a}_{\mu} (x) \nl 
& + & (2 \a')^{1/2} \xi^{\rho} (z_1) \xi^{\sigma} (z_2) \partial_{\rho} 
A_{\mu}^{a} (x) \partial_{\sigma} A_{\nu}^{a} (x) \Big] \Bigg>~~~. 
\label{c2''}
\eeqa
The expectation value can be computed simply using Wick theorem, according to
\beq
< X_\mu(z) X_\nu(w) > =  - g_{\mu \nu}\, {\cal G}(z,w)~~~,
\label{Greenfu}
\eeq
so that terms with an odd number of $\xi$ fields give a vanishing 
contribution. The result is
\beq
C^{(h)}_2(A) = - \frac{1}{4} 
{\tilde{F}}^{a}_{\mu \nu} (x) {\tilde{F}}^{a}_{\mu \nu}(x) 
(2 \a' )^2 \int dz_1 \int d z_2 ~\theta (z_1 - z_2 ) \partial_{z_1}
{\cal G}(z_1 , z_2 )  \partial_{z_2} {\cal G}(z_1 , z_2 )  
\label{c2final}
\eeq
where ${\tilde{F}}_{\mu \nu}^{a} = \partial_{\mu} A_{\nu}^{a} 
- \partial_{\nu} A_{\mu}^{a}$ is the abelian part of the field 
strength tensor.

Let us now turn to terms cubic in the external field. 
Here we use the fact that
\beq
{\rm Tr} \left[ A_{\mu} A_{\nu} A_{\rho}  \right]
= \frac{1}{4} \left[ {\rm i} f^{abc} + d^{abc} \right] 
A_{\mu}^a  A_{\nu}^{b}  A_{\rho}^c~~~.
\label{trace}
\eeq
The completely symmetric term does not give any contribution up to order 
$(2 \a' )^2 $, as one can easily see that the corresponding integrands are 
total derivatives. We are left with
\beqa
C^{(h)}_3(A) & = & \frac{{\rm i}}{4} ( 2 \a' )^2 f^{abc} \Bigg< \int 
\prod_{i=1}^{3} d z_i ~\partial_{z_1} \xi^{\mu} (z_1 ) 
\partial_{z_2} \xi^{\nu} (z_2 ) \partial_{z_3} \xi^{\rho} (z_3 ) \nl
& \times & \left\{ \xi^{\sigma} (z_3 ) A_{\mu}^{a} (x) A_{\nu}^{b} (x)
\partial_{\sigma} A_{\rho}^{c} (x)  \right. \label{c3} \\
& + & \xi^{\sigma} (z_2 )   A_{\mu}^{a} (x) \partial_{\sigma} 
A_{\nu}^{b} (x) A_{\rho}^{c} (x)  \nl
& + & \left. \xi^{\sigma} (z_1 )  \partial_{\sigma} A_{\mu}^{a} (x) 
A_{\nu}^{b} (x) A_{\rho}^{c} (x) \right\} \Bigg>~~~. 
\nonumber
\eeqa
Performing all possible contractions with \eq{Greenfu} we arrive at 
\beqa
C^{(h)}_3(A) & = &  \frac{{\rm i}}{8} 
( 2 \a' )^2 f^{abc} {\tilde{F}}^{a}_{\mu \nu} (x)
A_{\mu}^{b} (x) A_{\nu}^{c} (x) \nl
& \times & \int d z_1 \int d z_2 \int d z_3 ~\theta( z_1 - z_2 ) 
\theta( z_2 - z_3 )  \nl
& \times & \Bigg\{ \partial_{z_1} \partial_{z_2} {\cal G}( z_1 , z_2 ) \Big[
\partial_{z_3} {\cal G}(z_2 , z_3 ) - 
\partial_{z_3} {\cal G}( z_1 , z_3 )\Big] 
\label{c3''}\\
& + &     \partial_{z_1} \partial_{z_3} {\cal G}( z_1 , z_3 ) \Big[
\partial_{z_2} {\cal G}(z_1 , z_2 ) - \partial_{z_2} {\cal G}( z_2 , z_3 )  
\Big]  \nl
& + &  \partial_{z_2} \partial_{z_3} {\cal G}( z_2 , z_3 ) \Big[
\partial_{z_1} {\cal G}(z_1 , z_3 ) - \partial_{z_1} 
{\cal G}( z_1 , z_2 )\Big]
\Bigg\}~~~. \nonumber
\eeqa
This can be further simplified by partial integration of the double 
derivatives.
Discarding all terms with Green functions evaluated at pinching, and boundary
terms involving integrals around the entire boundary, we can reduce the
evaluation of $C^{(h)}_3(A)$ to the same integral that gives $C^{(h)}_2(A)$,
\beqa
C^{(h)}_3(A) & = &  \frac{{\rm i}}{2} 
( 2 \a' )^2 f^{abc} {\tilde{F}}^{a}_{\mu \nu} (x)
A_{\mu}^{b} (x) A_{\nu}^{c} (x) \label{c3final} \\
& \times & \int d z_1 \int d z_2 ~\theta( z_1 - z_2 ) 
\partial_{z_1} {\cal G}( z_1 , z_2 ) 
\partial_{z_2} {\cal G}( z_1 , z_2 )~~~,
\nonumber
\eeqa
which has the correct form to reconstruct the non-abelian field strength.
Similar manipulations can be performed on the term quartic in the external 
gauge fields. In terms of Green functions, it is given by
\beqa
C^{(h)}_4(A) & = & ( 2 \a '  )^2 {\rm Tr} 
\left( A_{\mu} (x)  A_{\nu} (x) A_{\rho} (x) 
A_{\sigma} (x)  \right)  \nl
& \times & \int  d z_1 \int dz_2 \int d z_3 \int d z_4  ~
\theta(z_1 - z_2 )
\theta ( z_2 - z_3) \theta ( z_3 - z_4 ) \nl 
& \times & \Big[ 
g^{\mu \nu} g^{\rho \sigma} \partial_{z_1} \partial_{z_2}
{\cal G}( z_1 , z_2 ) \partial_{z_3} \partial_{z_4}
{\cal G}( z_3 , z_4 ) \nl
& + &  g^{\mu \rho} g^{\nu \sigma} \partial_{z_1} \partial_{z_3}
{\cal G}( z_1 , z_3 ) \partial_{z_2} \partial_{z_4} {\cal G}( z_2 , z_4 ) 
\label{c4'}\\
& + &  g^{\mu \sigma} g^{\nu \rho} \partial_{z_1} \partial_{z_4}
{\cal G}( z_1 , z_4 ) \partial_{z_2} \partial_{z_3} {\cal G}( z_2 , z_3 ) 
\Big]~~~. \nonumber
\eeqa

Now we observe that the group factor for the middle term in the previous 
equation can be written as
\beq
{\rm Tr} ( A_{\mu} A_{\nu} A_{\mu} A_{\nu} ) = {\rm Tr} 
( A_{\mu} A_{\mu} A_{\nu} A_{\nu} )
- \frac{1}{4} f^{bcg} A_{\mu}^{b} A_{\nu}^{c} f^{dah} A_{\mu}^{d} A_{\nu}^{a}
\label{gfac}
\eeq
It is easy to see that only the last term in \eq{gfac} contributes 
to \eq{c4'}: in fact, in the sum of the other three terms 
of \eq{c4'} the integrand is completely symmetric in the exchange of 
any two punctures. The theta functions can thus be removed, 
and then the integrand is a total derivative. The only 
surviving contribution is
\beqa
C^{(h)}_4(A) & = & - \frac{1}{4} ( 2 \a '  )^2  f^{bcg} A_{\mu}^{b} (x) 
A_{\nu}^{c} (x) f^{dag} A_{\mu}^{d} (x) A_{\nu}^{a} (x) \label{c4''}\\
& \times & \int d z_1 \int d z_2 \int d z_3 \int d z_4~\theta (z_1 - z_2) 
\theta (z_2 - z_3 ) \theta ( z_3 - z_4 ) \nl  
& \times & \partial_{z_1} \partial_{z_3}
{\cal G}( z_1 , z_3 ) \partial_{z_2} \partial_{z_4} {\cal G}( z_2 , z_4 )~~~.
\nonumber  
\eeqa
Partial integration gives finally
\beqa
C^{(h)}_4(A) & = &  \frac{1}{4} ( 2 \a ' )^2  f^{bcg} A_{\mu}^{b} (x) 
A_{\nu}^{c} (x) f^{dag} A_{\mu}^{d} (x) A_{\nu}^{a} (x) \label{c4final}\\ 
& \times & \int d z_2 \int d z_3 ~\theta ( z_2 - z_3 ) \partial_{z_2} 
{\cal G}( z_2 , z_3 ) \partial_{z_3} {\cal G}( z_2 , z_3 )~~~. \nonumber
\eeqa
As expected, $C^{(h)}_2(A)$, $C^{(h)}_3(A)$ and $C^{(h)}_4(A)$ combine 
to reconstruct the non-abelian
field strength. Our results can be summarized in a very compact expression 
for the partition function defined by \eq{part}, valid for any 
number of string loops.
Omitting the vacuum contribution and higher orders in $\a'$, we get
\beq
Z_{P} ( A_{\mu} )= (2 \a ')^{2 -d/2} \int d^d x \left[- \frac{1}{4}
F_{\mu \nu}^a(x) F_{\mu \nu}^a(x) \right] \sum_{h=0}^{\infty} N^h g_{s}^{2h-2}
Z^{(h)} (d)~~~,
\label{partfin}
\eeq
where  
\beq
Z^{(h)} (d) = - g_{d}^{2} \int d {\cal{M}}_h 
\int dz_1 \int d z_2 ~ \theta (z_1 - z_2 ) \partial_{z_1}
{\cal G}(z_1 , z_2 )  \partial_{z_2} {\cal G}(z_1 , z_2 )~~~,
\label{esse}
\eeq
and
\beq
F_{\mu \nu}^{a} = \partial_{\mu} A_{\nu}^{a} - \partial_{\nu} A_{\mu}^{a}
+ g_d f^{abc} A_{\mu}^{b} A_{\nu}^{c}~~~.
\label{fmunu}
\eeq
Up to some constant prefactors, $Z^{(h)} (d)$ is the multiloop generalization
of the integral $Z(d)$ defined in \secn{effact}.

As we mentioned before, tree-level and one-loop contributions need extra care
because of the unfixed projective invariance of the functional integral. Here
we sketch how this problem can be handled, drawing on the results of the 
previous sections and on the ideas of \cite{tselett}. In the process,
as expected, we rederive the one-loop integral $Z(d)$.

At tree level, the projective infinity that would have been eliminated by 
fixing three of the punctures can be regularized by compactifying the 
integration region of the variables $z_i$, mapping them from the real axis 
to a circle. On a circle, following Ref.~\cite{tselett}, we can use the 
Green function 
\beqa
\hat G( \phi_1 , \phi_2 ) & = & \log \left[ 2{\rm i} \sin \left( 
\frac{\phi_1 - \phi_2 }{2} \right)\right]  \nl
& = & - \sum_{n=1}^{\infty} \frac{\cos n ( \phi_1 - \phi_2 )}{n}
+ \dots~~~.
\label{treegreen'}
\eeqa
The integrals over the punctures $\phi_i $ are now ordered in the interval 
$( 0, 2\pi )$. The dots in Eq. (\ref{treegreen'}) stand for terms 
independent of the punctures, that are irrelevant.

Using \eq{treegreen'}, we find that the basic integral appearing 
in $Z^{(0)}(d)$ is given by
\beqa
&  & \int_0^{2 \pi} d \phi_1 \int_0^{\phi_1} d \phi_2 ~\partial_{\phi_1}
{\hat G}(\phi_1 , \phi_2 )  \partial_{\phi_2} {\hat G}
(\phi_1 , \phi_2 ) \nl
& = & - \frac{2 \pi }{2} \int_{0}^{2 \pi} d \phi \sum_{n=1}^{\infty} 
\sin^2 n \phi = - \frac{(2 \pi)^2}{4} \sum_{n=1}^{\infty} 1 = 
\frac{(2 \pi)^2}{8}~~~,
\label{c2tse}
\eeqa
where we have regularized the divergent sum using \eq{zreg}.

Inserting this result in \eq{c2final} we get, at tree level,
\beq 
C_2^{(0)}(A) = - \frac{1}{4} {\tilde{F}}^{a}_{\mu \nu} (x) 
{\tilde{F}}^{a}_{\mu \nu}(x) (2 \a' )^2 \frac{(2 \pi)^2}{8}~~~.
\label{c2tree}
\eeq

If we compute  the six integrals in \eq{c3''}, and the three integrals in
\eq{c4'}, we see that the combination of $C_2^{(0)}(A)$, $C_3^{(0)}(A)$ and 
$C_4^{(0)}(A)$ in \eq{expansion} correctly reconstructs  
the full non-abelian gauge invariant action, in agreement with 
the results of Ref.~\cite{tselett}, where only $C_2^{(0)}(A)$ is explicitly 
computed while $C_3^{(0)}(A)$ and $C_4^{(0)}(A)$ are deduced from gauge 
invariance. The result is of the form
\beq
Z_{P}^{h=0} ( A_{\mu} )   =  - \frac{v}{4} ~\int d^d x  
\left[ F_{\mu \nu}^{a} (x) F_{\mu \nu}^{a} (x) \right]~~~,
\label{finh=0}
\eeq
where, with this regularization, $v = - 2 \pi^2$ in four dimensions. Notice
that $v$ turns out to equal the integration volume multiplied 
with $\zeta(0)$, and is clearly regularization-dependent.

The one-loop coefficients $C^{(1)}_i(A)$ can be similarly computed 
without fixing any of the punctures. We can however fix the location of the 
fixed points $\eta$ and $\xi$ of the Schottky group, as was done in the 
previous sections. Having done that, we are left with invariance under 
translations of the $z_i$, or, more conveniently, of the variables $\nu_i$, 
related to the $z_i$ by \eq{nui}. The difference with respect to the tree 
level integrals is that now the variables $\nu_i$ are integrated in an 
ordered way in the interval $(0 , \tau )$ instead of $ ( 0, 2 \pi )$. 
As a consequence, instead of the factor $2 \pi$ arising from the 
redundant integration of a translationally invariant function, 
we will get a factor $\tau$. We have to remember, however, to 
divide the computed expression by the volume of the translational part of 
the projective group, that has not been fixed, and that is equal to 
$ \int_{k}^{1} \frac{dz}{z} = - \log k = 2 \tau$. With the inclusion of this
factor the measure of integration over $k$ is
\beq
d {\cal{M}}_1 = \frac{dk}{k^2} \left[- \frac{1}{2 \pi} \log k \right]^{-d/2} 
\frac{1}{(- \log k)} \prod_{n=1}^{\infty} \left( 1- k^n \right)^{2-d}~~~.
\label{meash=1}
\eeq

The coefficients $C_2^{(1)}(A)$, $C_3^{(1)}(A)$ and $C_4^{(1)}(A)$ can be 
computed directly from Eqs. (\ref{c2final}), (\ref{c3''}) and (\ref{c4'}), 
so that no extra assumptions on the cancellation of boundary terms need 
to be made. We find, using the Green function given by \eq{newmtgreen},
\beq 
C_i^{(1)}(A) = - 2 C_i^{(0)}(A) \sum_{n=1}^{\infty} 
\left( \frac{1 + q^{2n}}{1- q^{2n}} \right)^2~~~,
\label{c2h=1}
\eeq
for $i = 2,3,4$. The one-loop contribution to the partition function 
defined by \eq{part} is thus given by
\beq
Z_{P}^{h=1} ( A_{\mu} )   =  \frac{N}{4} 
\frac{ g_{d}^{2}}{ (4 \pi)^{d/2}} (2 \a ')^{2 - d/2}\, Z(d) \, 
\int d^d x  \left[ - \frac{1}{4} F_{\mu \nu}^{a} (x)
F_{\mu \nu}^{a} (x) \right]~~~,
\label{finh=1}
\eeq
which is precisely the result of Eqs. (\ref{efflag}) and (\ref{loopefflag}),
with $Z(d)$ given in Eq. (\ref{Zint}).
We have thus verified that the general formalism developed in the first part 
of this section applies to the somewhat special cases $h = 0$ and $h = 1$, and
we have also verified that the assumptions made for general $h$, 
concerning the
cancellation of boundary terms and the regularization of the pinchings, are 
satisfied for $h < 2$. Notice finally that the present functional integral
derivation of the renormalization constants provides an alternative way for 
computing the normalization factor $C_h$ given in \eq{vertnorm}. The $h$-loop
contribution to $Z_P(A)$ contains in fact precisely a factor of 
$(2 \pi)^{(-d h)}$ corresponding to the integration over $h$ loop momenta,
a factor of $(2 \a')^{- d/2}$ from the change of measure given by \eq{measu},
and the correct power of the string coupling constant. 

\sect{Summary and perspectives}
\label{summa}

We have analyzed in detail three related methods to calculate renormalization 
constants in Yang-Mills theory, all based on a consistent prescription to
continue off the mass shell multigluon amplitudes derived from string
theory. Our prescription leads unambiguously to the background field method,
and is phrased directly in the language of dimensional regularization.

The use of the open bosonic string has introduced undesired difficulties due
to the presence of a tachyon, but we have shown explicitly how such 
difficulties can be overcome, at least at one loop. In this context, perhaps
the most significant observation is that divergences associated with the
tachyon are not regularized by analytic continuation of the space-time
dimension, whereas all divergences associated with Yang-Mills gauge bosons
turn into the usual poles at $d = 4$. On the other hand, using the simplest
available string theory has allowed us to phrase our technique directly
in the language of multiloop string amplitudes, while minimizing the
amount of string technology needed for the calculations. This may prove
necessary when practical multiloop calculations are attempted.

Our work complements what is known about string-derived on-shell multigluon
amplitudes~\cite{berkos}, and effective actions~\cite{mets}, in several ways.
Our prescription, clearly tied to, and in fact
inspired by the background field method in field theory, provides 
an off-shell extension of the analysis of Ref.~\cite{berdun}. 
By working with a simple string theory, and by not performing the partial
integration of double derivatives of the world-sheet Green functions that
appear in the amplitudes, we have been able to map explicitly each region
of moduli space that contributes to the field theory limit to a specific
set of Feynman diagrams. This mapping, which may prove useful in future 
applications going beyond one loop, has enabled us to identify 
precisely the regions of moduli space that contribute to the effective 
action, leading to considerable simplifications in the calculation of 
renormalization constants. At one loop, ultraviolet divergences of gluon
amplitudes turn out to be summarized by a single integral over the
string modular parameter, which still contains all contributions 
from the infinite tower of massive string states. The field theory limit 
of this integral is the gluon wave function renormalization of the 
background field method.
We have thus made a precise connection between the techniques of 
Ref.~\cite{berkos} and those of Ref.~\cite{mets}. In particular, we have
shown how the $\zeta$-function regularization of pinching singularities
introduced in Ref.~\cite{fratse1} is precisely suited to handle correctly
the contact terms left over by the propagation of tachyons in pinched
configurations. 
Finally, we have studied the partition function of an open string
interacting with a background Yang-Mills field, to all orders in string
perturbation theory, focusing on planar diagrams. As might have been
expected from the explicit results obtained at tree level and at one loop,
this partition function is a generating functional for the contributions
to the gluon effective action of string diagrams with an arbitrary number
of legs and loops. A formal analysis of this partition function shows
how string theory generates a gauge-invariant effective action for any
number of loops, and in fact leads to a multiloop generalization of the 
one-loop integral, $Z(d)$, generating the renormalization constants.
Tree-level and one-loop diagrams are special, because of the extra 
invariances that reduce the dimensionality of the corresponding string 
moduli space, but the differences can be handled and the known results are 
recovered also from the general $h$-loop formalism.

The general formalism developed in \secn{background} helps perhaps to shed
some light on the relationship between string theory and the background
field method. Examining the expression for the string partition function,
\eq{part}, one may observe that any string amplitude
may be written in terms of interactions with a background field, 
provided the bakground field satisfies the string equations of motion. In 
particular, ordinary gluon amplitudes correspond to a background
field constructed out of plane waves. The only apparent obstacle to this
interpretation is the presence in the full string amplitudes of the pinching
singularities, that correspond to interactions of the external fields
happening far away (on the string scale) from the string loops. This
is exactly the same problem that arises in field theory, in background 
field calculations of $S$-matrix elements. There, if one allows 
for interactions among the {\it background}
fields, one is forced to quantize them also, introducing a new gauge fixing 
term and constructing the corresponding propagator. Here, once the pinching
singularities are removed, the calculation of the effective action performed
along the lines of \secn{effact}, or of \secn{background}, can be simply
understood as an infrared limit of the calculation of a gluon scattering
amplitude, where a quasi-constant background field has taken the place 
of the plane wave representing the gluon.

A few remarks may be added concerning the choice of regularization scheme.
This paper makes extensive use of dimensional regularization, which is very 
practical if one whishes to compare the results with QCD calculation, where
it is by far the most commonly used scheme. However, it should be kept
in mind that this choice is not by any means the only possible one. In 
fact, it may not be the best choice if one wishes to extend to more than one 
loop the effective action calculations performed in \secn{effact}. In
general, $h$-loop contributions to renormalization constants appear
in dimensional regularization as coefficients of the single pole in
$\e = 2 - d/2$, and may be missed if one uses a method that neglects
$O(\e)$ corrections. Regularization with, say, a momentum space cutoff
may be more appropriate, since in that case divergences are further
classified according to the power of the cutoff involved (see for
example Ref.~\cite{mets}). It may also be argued that string theory
by itself provides a regularized version of Yang-Mills theory, with the 
role of the cutoff being played by
$1/\a'$.  In fact, this ``stringy'' regularization is quite similar in
spirit to Pauli-Villars regularization, in the sense that the theory is
made finite by adding an infinite number of new degrees of
freedom with heavy masses and with carefully selected coefficients.
It might be instructive to study how this ``stringy'' regularization
translates in a field theory language.

Summarizing, we believe we have shown that the multiloop string technology 
needed to study perturbative Yang-Mills theory exists, and the complexity
of the necessary calculations may remain manageable, provided one uses
a sufficiently simple string theory, as we have done in this paper.
This may have wide-ranging applications, not only to practical calculations
of phenomenological interest, but more generally to develop a deeper
understanding of the organization of the perturbative expansion of
gauge theories. String theory has proven to be remarkably clever in
organizing gauge theory amplitudes both at tree level and at one loop,
suggesting the best choices of color and helicity organizations, as well
as the best combination of gauges. It is fair to hope that similar results
may be achieved at least at two loops. On the other hand, the very 
existence of explicit all-order expressions for the amplitudes in string 
theory suggests that these tools might also be useful to derive 
all-order results in field theory, as was mentioned in \secn{twopoint}.
A simple example of such a derivation is given by
our discussion of string propagation in a background field.

\vskip 2.5cm

{\large {\bf {Acknowledgements}}}
\vskip 0.5cm
One of us (R. M.) would like to thank NORDITA for the kind hospitality
during the completion of this work. L. M. thanks G. Marchesini for a 
stimulating remark and D. Dunbar for instructive conversations.
This research was partially supported by the EU, within the framework of 
the program ``Gauge Theories, Applied Supersymmetry and Quantum Gravity'', 
under contract SCI-CT92-0789.

\vskip2cm

\appendix{\Large {\bf {Appendix A}}}
\label{appa}
\vskip 0.5cm
\renewcommand{\theequation}{A.\arabic{equation}}
\setcounter{equation}{0}

In this appendix we compute the normalization coefficients of the 
$h$-loop string amplitudes, $C_h$, and of the gluon states,
${\cal{N}}_0$, given respectively in
Eqs. (\ref{vertnorm}) and (\ref{glunorm}). Our strategy to do so is the
following: we first compute the three and four-gluon amplitudes
at tree level, and require that in the limit $\alpha'\to 0$ they reduce 
to the corresponding amplitudes derived from the Yang-Mills Lagrangian. 
This fixes $C_0$ and ${\cal{N}}_0$ as functions of the Yang-Mills coupling 
constant $g_d$; we then check that these normalizations are consistent 
with the factorization properties required by unitarity, 
and finally we implement the sewing procedure to obtain
the multiloop amplitudes, and their overall normalization $C_h$. 

As discussed in section 4, in the open bosonic string
the color-ordered three-gluon amplitude at tree level is
\bea
A^{(0)}(p_1,p_2,p_3) & = &  C_0 \, {{\cal{N}}_0}^3 
\,{\rm Tr}(\lambda^{a_1}\lambda^{a_2}\lambda^{a_3})~
\sqrt{2\alpha'}\,
\Big(-\ve_1\cdot\ve_2\,p_2\cdot\ve_3 \nonumber \\
& & - \, \ve_2\cdot\ve_3\,p_3\cdot\ve_1
- \ve_3\cdot\ve_1\,p_1\cdot\ve_2 +
O(\a') \Big)~~~,
\label{athree}
\ena
where $\ve_i$, $p_i$, $a_i$ are the polarization, the momentum 
and the color index of the $i$-th gluon, and the $\lambda$'s are the 
generators of $SU(N)$ in the fundamental representation. For them we
take the standard field theory conventions, {\it i.e.}
\beq
{\rm Tr}(\lambda^a\lambda^b) = \frac{1}{2}~\delta^{ab}~~~,~~~
{\rm Tr}(\lambda^a\,[\lambda^b\, , \,\lambda^c]) = \frac{{\rm i}}{2}~
f^{abc}~~~,
\label{agennorm}
\eeq
where $f^{abc}$ are the group structure constants. We notice that
this choice implies in particular that
\bea
\sum_{a} {\rm Tr}(A\,\lambda^{a})\,{\rm Tr}(\lambda^{a}B) & = &
\frac{1}{2}\, {\rm Tr}(A\,B) -\frac{1}{2N}\,{\rm Tr}(A)\,
{\rm Tr}(B)~~~, \label{trtr} \\
\sum_{a} {\rm Tr}(\lambda^{a}A\,\lambda^{a}) & = &
\frac{N}{2} \,{\rm Tr}(A) -\frac{1}{2N}\,{\rm Tr}(A)~~~,
\label{tr}
\ena
with $A$ and $B$ being arbitrary $N \times N$ matrices.

With this normalization of the color matrices, it is easy to check that 
in the limit $\alpha'\to 0$, \eq{athree} reduces to the color-ordered 
three-gluon amplitude derived from the Yang-Mills 
Lagrangian if
\beq
C_0\,{{\cal{N}}_0}^3 = \frac{4\,g_d}{\sqrt{2\alpha'}}~~~,
\label{c03}
\eeq
where $g_d$ is the (dimensionful) Yang-Mills coupling constant in $d$ 
dimensions.

The color-ordered four-gluon amplitude at tree level is
\bea
A^{(0)}(p_1,p_2,p_3,p_4) & = &  \, C_0 \, {{\cal{N}}_0}^4\,
{\rm Tr}(\lambda^{a_1}\lambda^{a_2}\lambda^{a_3}\lambda^{a_4})~
\frac{\Gamma(1-\alpha's)\Gamma(1-\alpha't)}{\Gamma(1+\alpha'u)}
\nonumber \\
& & \Bigg[ \, \frac{u}{s(1+\a's)}\,
\ve_1\cdot\ve_2\,\ve_3\cdot\ve_4 
+\frac{1}{1+\a'u}
\ve_1\cdot\ve_3\,\ve_2\cdot\ve_4
\nonumber \\ 
& & \, \, + \, \, \frac{u}{t(1+\a't)}
\,\ve_1\cdot\ve_4\,\ve_2\cdot\ve_3 
\nonumber \\
& & \, \, - \, \frac{2}{t}\,\Big(
\ve_1\cdot\ve_3\,\ve_4\cdot p_1
\,\ve_2\cdot p_3 +
\ve_1\cdot\ve_4\,\ve_3\cdot p_1
\,\ve_2\cdot p_4 
\nonumber \\
& &~~~~~~+ \,
\ve_2\cdot\ve_3\,\ve_1\cdot p_3
\,\ve_4\cdot p_2 +
\ve_2\cdot\ve_4\,\ve_3\cdot p_2
\,\ve_1\cdot p_4 \Big)
\nonumber \\
& & \, \, - \, \frac{2}{s}\,\Big(
\ve_1\cdot\ve_2\,\ve_4\cdot p_2
\,\ve_3\cdot p_1 +
\ve_1\cdot\ve_3\,\ve_4\cdot p_3
\,\ve_2\cdot p_1 
\nonumber \\
& &~~~~~~+ \,
\ve_2\cdot\ve_4\,\ve_3\cdot p_4
\,\ve_1\cdot p_2 +
\ve_3\cdot\ve_4\,\ve_1\cdot p_3
\,\ve_2\cdot p_4 \Big)
\nonumber \\
& & \, \, - \, \frac{2\,u}{s\,t}\,\Big(
\ve_1\cdot\ve_2\,\ve_4\cdot p_1
\,\ve_3\cdot p_2 +
\ve_1\cdot\ve_4\,\ve_3\cdot p_4
\,\ve_2\cdot p_1 
\nonumber \\
& &~~~~~~+ \,
\ve_2\cdot\ve_3\,\ve_1\cdot p_2
\,\ve_4\cdot p_3 +
\ve_3\cdot\ve_4\,\ve_2\cdot p_3
\,\ve_1\cdot p_4 \Big)
\nonumber \\
& & \, \, + \, \, O(\a') \, \Bigg]~~~,
\label{afour}
\ena
where, as usual, $s=-(p_1+p_2)^2$, $t=-(p_1+p_4)^2$ and 
$u=-(p_1+p_3)^2$.
In the limit $\a'\to 0$ this expression reduces to the color-ordered
four-gluon amplitude (consisting of both 1PI and exchange diagrams in 
the $s$ and $t$ channels) derived from the Yang-Mills lagrangian
if
\beq
C_0\,{{\cal{N}}_0}^4 = 4\,g_d^2~~~.
\label{c04}
\eeq
Combining \eq{c03} and \eq{c04}, we easily derive
\bea
{\cal N}_0&=& g_d\,\sqrt{2\a'}~~~,
\label{n0} \\
C_0 &=& \frac{1}{\left(g_d\,\a'\right)^2}~~~.
\label{c0}
\ena

We now verify that these normalizations are consistent with the 
factorization properties required by unitarity. To do so, let us 
consider the amplitude given in \eq{afour}, which has poles at
\beq
s=m_I^2\equiv\frac{I}{\a'}~~~~~{\rm or}~~~~~t=m_I^2~~~,
\label{spoles}
\eeq 
for $I=-1,0,1, \ldots$.
Each pole is associated with the exchange of an intermediate
string state $|I\rangle$ of mass $m_I$ in the $s$ or in the 
$t$ channel ($I=-1$ corresponding to tachyon exchange, $I=0$ to gluon 
exchange, and so forth).
Unitarity requires that the residues of these poles have certain simple
factorization properties. As far as the $s$ channel is concerned, these
can be symbolically represented by saying that, for $s\to m_I^2$,
\beq
A^{(0)}(p_1,p_2,p_3,p_4) \sim {\sum}_{(I)}
A^{(0)}_I(p_1,p_2,q)\,\frac{1}{q^2+m_I^2}\,A^{(0)}_I(-q,p_3,p_4)~~~.
\label{factor}
\eeq
In this formula $q=-p_1-p_2=p_3+p_4$ is the exchanged momentum, 
while $A^{(0)}_I(p_1,p_2,q)$ denotes the ordered amplitude 
involving two gluons with momenta $p_1$ and $p_2$, 
and the $I$-th string state with momentum $q$. 
The symbol ${\sum}_{(I)}$ stands for a sum over all these 
intermediate states, including for each one of them a sum over all 
its quantum numbers.

\eq{factor} is an example of the well-known fact that a four-string
amplitude can be interpreted as a sewing
of two three string amplitudes, with $1/(q^2+m_I^2)$ acting as a
propagator for the $I$-th state $|I\rangle$. Notice that
\beq
\frac{1}{q^2+m_I^2} ~|I\rangle = \frac{\a'}{L_0-1} ~|I\rangle~~~,
\label{prop}
\eeq
where $L_0$ is the zero mode of the Virasoro algebra.
This equation allows to identify $\a'/(L_0-1)$ as the sewing operator
\footnote{We refer to \cite{DFLS} for a thorough discussion of
the sewing procedure and for an analysis of different forms
of sewing operators, including their expressions when the
ghost degrees of freedom are taken into account.}.
Sice we have been concentrating on planar contributions,
in the sewing process we must be 
careful to preserve the ordering of the color indices,  
{\it i.e.} the sewing must also occur in a planar way.
For $SU(N)$ this is realized by retaining in the combinations of the 
group color factors only those terms which are leading in the formal 
limit $N\to\infty$. 

To check whether \eq{factor} is satisfied or not, we first 
examine the gluon exchange ($I=0$). From 
\eq{afour} we easily see that when $s\to 0$
\bea
A^{(0)}(p_1,p_2,p_3,p_4) & \sim &  2\, C_0 \, {{\cal{N}}_0}^4\,
{\rm Tr}(\lambda^{a_1}\lambda^{a_2}\lambda^{a_3}\lambda^{a_4})~
\frac{1}{s}
\nonumber \\
& \times & \!\!
\Bigg[-\frac{t}{2}\,
\ve_1\cdot\ve_2\,\ve_3\cdot\ve_4 
-\ve_1\cdot\ve_2\,\ve_4\cdot p_2
\,\ve_3\cdot p_1 
-\ve_1\cdot\ve_3\,\ve_4\cdot p_3
\,\ve_2\cdot p_1 
\nonumber \\
&&\!\!\!\!\!\!\!\!
-\ve_2\cdot\ve_4\,\ve_3\cdot p_4
\,\ve_1\cdot p_2 
-\ve_3\cdot\ve_4\,\ve_1\cdot p_3
\,\ve_2\cdot p_4 
+\ve_1\cdot\ve_2\,\ve_4\cdot p_1
\,\ve_3\cdot p_2 
\nonumber \\
&&\!\!\!\!\!\!\!\!
+\ve_1\cdot\ve_4\,\ve_3\cdot p_4
\,\ve_2\cdot p_1 
+\ve_2\cdot\ve_3\,\ve_1\cdot p_2
\,\ve_4\cdot p_3 
+\ve_3\cdot\ve_4\,\ve_2\cdot p_3
\,\ve_1\cdot p_4 
\nonumber \\
&&\!\!\!\!\!\!\!\!
+~~O(\a')\Bigg]~~~.
\label{resglu}
\ena
On the other hand, using \eq{athree} for $A^{(0)}_0$ 
and writing explicitly the sum over the intermediate gluons
as a sum over colors and polarizations, we have
\bea
& & {\sum}_{(0)}
A^{(0)}_0(p_1,p_2,q)\,\frac{1}{q^2}\,A^{(0)}_0(-q,p_3,p_4)
\label{resglufact} \\
& = & \frac{C_0^2 {\cal N}_0^6 2 \a'}{q^2}~\sum_{b} 
\Big({\rm Tr}(\lambda^{a_1}\lambda^{a_2}\lambda^{b})
{\rm Tr}(\lambda^{b}\lambda^{a_3}\lambda^{a_4})\Big)
\nonumber \\
& \times & \sum_{\ve}\Bigg[
\Big(\ve_1\cdot\ve_2\,p_2\cdot\ve +
\ve_2\cdot\ve\,q\cdot\ve_1
+ \ve\cdot\ve_1\,p_1\cdot\ve_2 +
O(\a') \Big)\nonumber \\
& & \Big(\ve\cdot\ve_3\,p_3\cdot\ve_4
+ \ve_3\cdot\ve_4\,p_4\cdot\ve
-\ve_4\cdot\ve\,q\cdot\ve_3 +
O(\a') \Big)\Bigg]~~~. \nonumber
\ena
We now use the identity
\beq
\sum_{\ve} a\cdot\ve\,b\cdot\ve = a\cdot b~~~,
\label{polsum}
\eeq
and perform the sum over colors using \eq{trtr}, and keeping only the 
planar contribution. 
After some straightforward algebra,
it is easy to see that \eq{resglu} and \eq{resglufact}
coincide if
\beq
C_0\,{{\cal N}_0}^2\,\a'=2~~~,
\label{c0n0}
\eeq
which is satisfied by ${\cal N}_0$ and $C_0$, as given in Eqs. (\ref{n0}) 
and (\ref{c0}).

It is instructive to examine also the contributions due to exchanges
of other string states. For example, let us consider the tachyon 
($I=-1$). In order to check \eq{factor} in this case, 
we need to know the ordered amplitude $A^{(0)}_{-1}$ involving two 
gluons and one tachyon. This can be easily computed in the
operator formalism, and one finds
\beq
A^{(0)}_{-1}(p_1,p_2,q) = C_0\,{{\cal N}_0}^2\,{\cal N}_{-1}\,
{\rm Tr}(\lambda^{a_1}\lambda^{a_2}\lambda^{b})\,
\left(\ve_1\cdot\ve_2+O(\a')\right)~~~,
\label{glutac}
\eeq
where ${\cal N}_{-1}$ denotes the normalization of the tachyon state.
Repeating the same steps as before, we can see that
the residue of the four-gluon amplitude at the tachyon pole, \eq{afour},
factorizes into the product of two three-point amplitudes, like \eq{glutac},
if $C_0\,{{\cal N}_{-1}}^2\,\a'=2$.
Without much difficulty it is possible to generalize this relation to 
the $I$-th string state, with normalization ${\cal N}_I$, 
and thus one can conclude that unitarity implies that
\beq
C_0\,{{\cal N}_{I}}^2\,\a'=2
\label{c0ni}
\eeq
for all $I$.

Let us now turn to the normalization $C_1$ of the one-loop amplitudes.
Since a loop can be obtained by sewing two legs of a tree-level
amplitude, the coefficient $C_1$ can be deduced from $C_0$. To see this,
let us consider a color-ordered amplitude for the scattering of $M$ gluons 
and two arbitrary states $I$ and $J$ (which for simplicity
we take to be adjacent in the chosen ordering along the loop), at tree level. 
We schematically denote such amplitude by
\beq
A^{(0)}_{IJ}(p_I, p_J) ~=~ 
C_0\,{\cal N}_I\,{{\cal N}_0}^M\,{\cal N}_J\,
{\rm Tr}(\lambda^{b_I}\lambda^{a_1}\cdots\lambda^{a_M}\lambda^{b_J})~
K_{IJ}(p_I, p_J)~~~,
\label{ij}
\eeq
where $K_{IJ}(p_I,p_J)$ is an appropriate kinematic factor.
To obtain the $M$-gluon amplitude at one loop we first identify
the states $I$ and $J$ by means of the sewing operator, which
carries a factor of $\a'$, and then we integrate over their momentum
$q$ and sum over their color index $b$. Finally, we must sum over all 
possible intermediate states. Thus, upon sewing, \eq{ij} is replaced
by
\beq
A^{(1)}(p_1, \ldots p_M) ~=~ 
\sum_{J}\,C_0\,{{\cal N}_J}^2\,{{\cal N}_0}^M\,\sum_b
{\rm Tr}(\lambda^{b}\lambda^{a_1}\cdots\lambda^{a_M}\lambda^{b})\,
\int \frac{d^dq}{(2\pi)^d}~{\tilde K}_{JJ}(q,q)~~~,
\label{ii}
\eeq
where ${\tilde K}_{JJ}(q,q)$ is what it is obtained from $K_{JJ}(q,q)$
with the inclusion of the sewing operator.
The integral over $q$ produces a factor of $(2\a')^{-d/2}$ and leaves
the non-zero mode part of ${\tilde K}_{JJ}(q,q)$, which we denote by
${\tilde K}'_{JJ}$, whereas the sum over colors yields
\beq
\sum_b
{\rm Tr}(\lambda^{b}\lambda^{a_1}\cdots\lambda^{a_M}\lambda^{b})\, = \,
\frac{N}{2}~{\rm Tr}(\lambda^{a_1}\cdots\lambda^{a_M})~~~,
\nonumber 
\eeq
provided we use \eq{tr}, and keep only the planar contribution.
Then, using \eq{c0ni}, \eq{ii} becomes
\beq
A^{(1)}_P(p_1, \ldots p_M) ~=~ 
\frac{1}{(2\pi)^d}\,(2\a')^{-{d}/{2}}\,{{\cal N}_0}^M\,
N\,{\rm Tr}(\lambda^{a_1}\cdots\lambda^{a_M})\,\sum_{J}
{\tilde K}'_{JJ}~~~.
\label{ii'}
\eeq
This is the schematic expression of the $M$-gluon planar amplitude
at one loop, and from it we can deduce that
\beq
C_1 = \frac{1}{(2\pi)^d}\,(2\a')^{-{d}/{2}}~~~.
\label{c1}
\eeq

When we iterate this sewing procedure to generate more loops,
we realize that 
\beq
\frac{C_h}{C_{h-1}}=
\frac{1}{(2\pi)^d}\,(2\a')^{-{d}/{2}}\,g_d^2\,{\a'}^2~~~,
\nonumber
\eeq
which leads to
\beq
C_h=\frac{1}{(2\pi)^{dh}}\,(2\a')^{-{dh}/{2}}\,
\,g_d^{2h-2}\,(\a')^{2h-2}~~~.
\label{ch}
\eeq
If we introduce the dimensionless string coupling constant
\beq
g_s = \frac{g_d}{2}\,(2\a')^{1-d/4}~~~,
\label{gs}
\eeq
it is possible to rewrite \eq{ch} as
\beq
C_h= \frac{1}{(2\pi)^{dh}}\,{g_s}^{2h-2}\,\frac{1}{(2\a')^{d/2}}~~~.
\label{chf}
\eeq
All factors in this expression have a simple interpretation: the first
ensures the canonical normalization of all loop momentum integrals,
the second contains the correct power of the string coupling costant,
dictated by string perturbation theory, and finally the last is necessary
to give the vacuum amplitude the proper dimension, $({\rm length})^{-d}$.

\vskip 2cm

\appendix{\Large {\bf {Appendix B}}}
\label{appb}
\vskip 0.5cm
\renewcommand{\theequation}{B.\arabic{equation}}
\setcounter{equation}{0}
 
In this appendix we compute the integral $Z(d)$, \eq{Zint}, in the field 
theory limit, and thus derive \eq{divefflag} from \eq{loopefflag}.
 
The first important step is the observation that the sum appearing in 
$Z(d)$ can be written as
\beq
\sum_{n=1}^{\infty} \left( \frac{1+ q^{2n}}{1 - q^{2n}} \right)^2 = 
\sum_{n=1}^{\infty} \left[ 1 + \frac{4q^{2n}}{(1- q^{2n})^2} \right] =
-2q \frac{d}{dq} \log \left[q^{1/4} \prod_{n=1}^{\infty}
( 1 - q^{2n})\right]~~,
\label{sumid}
\eeq
where, following Ref.~\cite{mets}, we have regularized the divergent sum 
$\sum\limits_{n=1}^\infty 1$ by means of \eq{zreg}.

It is interesting then to notice that the result of the integration over the 
punctures in the open string becomes very similar to what has been found by 
Kaplunov-ski~\cite{kaplu} for the closed heterotic string.
 
Using the identity
\beq
f(q^2) \equiv \prod_{n=1}^{\infty} (1 - q^{2n}) = 
\left( - \frac{2 \pi}{\log k}\right)^{-1/2} k^{1/24} 
{\rm e}^{- \pi^2 /(6 \log k)} f(k)~~~,
\label{modid}
\eeq
with $k \equiv {\rm e}^{-2 \tau}$ as usual, we can write $Z(d)$ as
\beqa
Z(d) & = & 16 \int_{0}^{\infty} {\cal D} \tau \, \tau \, k \,
\frac{d}{dk} \log \left[ \left( - \frac{ 2 \pi}{\log k}\right)^{-1/2} 
k^{1/24} \, {\rm e}^{- \pi^2 /(3 \log k)} f(k) \right]  \nl
& = & 16 \int_{0}^{\infty} {\cal D} \tau 
\left\{ - \frac{1}{4} - \frac{\pi^2}{12 \tau}
+ \frac{\tau}{24} - \tau \sum_{n=1}^\infty \frac{n k^n}{1 - k^n}
\right\}~~~.
\label{calcZ}
\eeqa
By expanding the partition function
\beq
\left[ f(k) \right]^{2 -d} = 1 + (d-2) k + O(k^2)~~~,
\label{partfunc}
\eeq
we see that the coefficient of the term $\tau^{1-d/2}$ in the integral 
contains an expansion in powers of $k$, possibly also with some extra factors 
coming from the first two terms in braces in \eq{calcZ}.

The leading term, $O(1/k)$, corresponds to tachyon exchange in the 
loop and, as always, we just neglect it by hand. We neglect also 
terms $O(\frac{1}{k (\log k)^a})$, where $a=1,2$, for the same reason.

The next term is of $O(1)$ in $k$, and corresponds to the exchange of massless 
gluons. It is given by
\beq
\frac{2}{3} (d - 26) \int_{0}^{\infty} d \tau~\tau^{1 - d/2}
{\rm e}^{- 2 \a' m^2 \tau} = \frac{2}{3} (d - 26) (2 \a' m^2)^{d/2 - 2} 
\, \Gamma \left(2 - \frac{d}{2} \right)~~~,
\label{maylast}
\eeq
where a term with an infrared cutoff $m^2$, reminiscent of
the exponentials of Green functions that we discarded, has been added. 
We stress again that the field theory 
limit is performed by sending $\a'$ 
to zero while keeping the ultraviolet cutoff $\e$ fixed and positive.
 
\eq{divefflag} is obtained by expanding for small $\e$ after inserting
\eq{maylast} in \eq{loopefflag}.
 
Higher order terms in \eq{calcZ}, $O(k^{I+1})$ 
with $ I>0$, are negligible 
in the field theory limit. This is clearly understood in string theory
remembering that the power $I$ is related to 
the mass of the string excitation 
exchanged in the loop, and to $\a'$, by the 
formula $I = \a' m_{I}^{2}$ 
(see also \eq{spoles}).


\vskip 2cm


\begin{thebibliography}{99}

\bibitem{copgroup} See, for example, P. Di Vecchia, {\it ``Multiloop 
amplitudes in string theory''} in Erice, {\it Theor. Phys.} (1992), 
and references therein. 
 
\bibitem{mets} R. R. Metsaev and A. A. Tseytlin,
{\it Nucl. Phys.} {\bf B 298} (1988) p. 109.

\bibitem{mina} J. A. Minahan, {\it Nucl. Phys.} {\bf B 298} (1988) p. 36.

\bibitem{tayven} T. R. Taylor and G. Veneziano, {\it Phys. Lett.} 
{\bf 212 B} (1988) p. 147.

\bibitem{kaplu} V. S. Kaplunovsky, {\it Nucl. Phys.} {\bf B 307}
(1988) p. 145, and {\it Nucl. Phys.} {\bf B 382} (1992) 
p. 436, hep-th/9205070.

\bibitem{berkosbet} Z. Bern and D. A. Kosower, {\it Phys. Rev.}
{\bf D 38} (1988) p. 1888.

\bibitem{berkoswfr} Z. Bern and D. A. Kosower, {\it Nucl. Phys.}
{\bf B 321} (1989) p. 605.

\bibitem{berkos} Z. Bern and D. A. Kosower, {\it Nucl. Phys.}
{\bf B 379} (1992) p. 451. 

\bibitem{fiveglu} Z. Bern, L. Dixon and D. A. Kosower,
{\it Phys. Rev. Lett.} {\bf 70} (1993) p. 2677, hep-ph/9302280.

\bibitem{berbos} Z. Bern, {\it Phys. Lett.} {\bf 296 B} (1992)
p. 85.

\bibitem{berdun} Z. Bern and D. C. Dunbar, {\it Nucl. Phys.}
{\bf B 379} (1992) p. 562.

\bibitem{strass} M. J. Strassler, {\it Nucl. Phys.} {\bf B 385} (1992)
p. 145, hep-ph/9205205.

\bibitem{schu} M. G. Schmidt and C. Schubert, {\it Phys. Lett.}
{\bf 331 B} (1994) p. 69, hep-th/9403158.

\bibitem{kaj} K. Roland, {\it Phys. Lett.} {\bf 289 B} (1992) p. 148.

\bibitem{letter} P. Di Vecchia, A. Lerda, L. Magnea and R. Marotta,
{\it Phys. Lett.} {\bf 351 B} (1995) p. 445, hep-th/9502156. 

\bibitem{fratse1} E. S. Fradkin and A. A. Tseytlin, {\it Phys. Lett.} 
{\bf 163 B} (1985) p. 123.

\bibitem{scho} P. Di Vecchia, F. Pezzella, M. Frau, K. Hornfeck, A. Lerda and 
S. Sciuto, {\it Nucl. Phys.} {\bf B 322} (1989) p. 317. 

\bibitem{DFLS} P. Di Vecchia, M. Frau, A. Lerda and S. Sciuto, 
{\it Nucl. Phys.} {\bf B 298} (1988) p. 526.

\bibitem{GSW} M. B. Green, J. H. Schwarz and E. Witten, 
{\it ``Superstring Theory''}, Cambridge University Press (1987).

\bibitem{berkosrol} Z. Bern, D. A. Kosower and K. Roland,
{\it Nucl. Phys.} {\bf B 334} (1990) p. 309.

\bibitem{abbott} L. F. Abbott, {\it Nucl. Phys.} {\bf B 185} 
(1981) p. 189.

\bibitem{abbgrisch} L. F. Abbott, M. T. Grisaru and R. K. Schaefer,
{\it Nucl. Phys.} {\bf B 229} (1983) p. 372.

\bibitem{tselett} A.A. Tseytlin, {\it Phys. Lett.} {\bf 202 B} (1988) p. 81.

\end{thebibliography}
\end{document}